\documentclass[12pt,preprint]{aastex}

\shorttitle{A comparison of methods for estimating CMEs kinematics}
\shortauthors{Mishra et al.}
\usepackage{pdflscape}
\usepackage{rotating}

\begin{document}

\title{A comparison of reconstruction methods for the estimation of CME kinematics based on SECCHI/HI observations}

\author{Wageesh Mishra\altaffilmark{1}, Nandita Srivastava\altaffilmark{1}, Jackie A. Davies\altaffilmark{2}}
\affil{Udaipur Solar Observatory, Physical Research Laboratory, P. O. Box 198, Badi Road,
Udaipur 313001, India}
\email{wageesh@prl.res.in}

\altaffiltext{1}{Udaipur Solar Observatory, Physical Research Laboratory, P. O. Box 198, Badi Road,
Udaipur 313001, India}

\altaffiltext{2}{RAL Space, Rutherford Appleton Laboratory, Harwell Oxford, OX11 0QX, UK}

\begin{abstract}
A study of the kinematics and arrival times of CMEs at Earth, derived from time-elongation maps (J-maps) constructed from STEREO/Heliospheric Imager (HI) observations, provides an opportunity  to understand the heliospheric evolution of CMEs in general. We implement various reconstruction techniques, based on the use of time-elongation profiles of propagating CMEs viewed from single or multiple vantage points, to estimate the dynamics of three geo-effective CMEs. We use the kinematic properties, derived from analysis of the elongation profiles, as inputs to the Drag Based Model for the distance beyond which the CMEs cannot be tracked unambiguously in the J-maps. The ambient solar wind into which these CMEs, which travel with different speeds, are launched, is different. Therefore, these CMEs will evolve differently throughout their journey from the Sun to 1 AU. We associate the CMEs, identified and tracked in the J-maps, with signatures observed in situ near 1 AU by the WIND spacecraft. By deriving the kinematic properties of each CME, using a variety of existing methods, we assess the relative performance of each method for the purpose of space weather forecasting. We discuss the limitations of each method, and identify the major constraints in predicting the arrival time of CMEs near 1 AU using heliospheric imager observations.

\end{abstract}


\section{Introduction}

Coronal Mass Ejections (CMEs) are episodic expulsions of magnetised plasma from the Sun into the heliosphere, which are responsible for about 85 $\%$ of intense geomagnetic storms \citep{Zhang2007, Echer2008}. Since the discovery of CMEs by the OSO-7 orbiting coronagraph \citep{Tousey1973}, thousands of CMEs have been observed using a series of coronagraphs \citep{Gosling1974,Sheeley1980, MacQueen1980,Fisher1981,Brueckner1995,Howard2008}. Before the era of heliospheric imagers, CMEs near the Sun were mainly observed using space based coronagraphs, and near the Earth using in situ instruments.  Using the Large Angle Spectrometric Coronagraph (LASCO) on board the SOlar and Heliospheric Observatory (SOHO) \citep{Brueckner1995}, studies of the dynamics of CMEs and estimates of their arrival time at 1 AU have been performed by various authors \citep{Gopalswamy2000, Gopalswamy2005, Michalek2004, Schwenn2005}. CMEs have been observed at larger distances from the Sun using interplanetary scintillation \citep{Hewish1964} and the zodiacal light photometer on-board the twin Helios spacecraft \citep{Richter1982}. With the launch of the Solar Mass Ejection Imager (SMEI: \citealt{Eyles2003}) on-board the Coriolis spacecraft in 2003, tracking of interplanetary disturbances using white light observations became possible over almost the entire sky. SMEI was switched off in 2011 and now only the Heliospheric Imagers (HIs:  \citealt{Eyles2009}), launched on-board the twin Solar TErrestrial RElations Observatory (STEREO) mission \citep{Kaiser2008} in 2006, have the capability to continuously image CMEs out to 1 AU and beyond, although with a field of view limited to around the ecliptic.  

The twin STEREO spacecraft move ahead of and behind the Earth in its orbit with their angular separation increasing by 45$\arcdeg$ per year. The STEREO observations enable us to perform 3D reconstruction of selected features of a CME based on suitable assumptions, some of which can be quite simple. The Sun-Earth Connection Coronal and Heliospheric Investigation (SECCHI: \citealt{Howard2008}) suite on board each STEREO spacecraft comprises an extreme ultraviolet imager (EUVI: 1-1.7 R$_\sun$ in the plane of the sky, POS), two coronagraphs (COR1:  1.4-4.0 R$_\sun$ POS and COR2: 2.5-15.0 R$_\sun$ POS) and two Heliospheric Imagers (HI1 and HI2). The COR1 and COR2 fields of view (FOVs) are centered on the Sun (0$\arcdeg$ elongation). HI1 has a 20$\arcdeg$ FOV, with its boresight centred at 14$\arcdeg$ elongation; HI2 has a wider FOV of 70$\arcdeg$  and its boresight is aligned at 53.7$\arcdeg$ elongation.

Since heliospheric imagers view CMEs via their Thomson scattered signal integrated along each line-of-sight, direct determination of the position of CME features from these observations is not possible. However, a number of techniques have been developed, based on a variety of assumptions regarding their geometry, propagation direction and speed, to derive CME kinematics by exploiting SMEI and HI observations (e.g. \citealt{Howard2006, Jackson2006, Jackson2007, Jackson2008, Kahler2007, Sheeley2008,Howard2009, Tappin2009,Lugaz2009, Lugaz2010, Liu2010, Lugaz2010.apj,Davies2012, Davies2013}). These techniques are based on a different set of assumptions which make them independent of each other to some degree. As will be explained further in next section,  several  of the methods treat CMEs as a point, while other methods consider CMEs to have a larger-scale geometry. In some of the methods, CMEs are assumed to  propagate with a constant speed while other methods can provide an estimate of the time variations of a CME’s speed as it propagates through the heliosphere. Since  some of the key questions in CME and space weather research relate to the propagation of CMEs, we consider it an obvious next step to attempt to ascertain the relative merits of these various reconstruction methods for estimating the kinematic properties of CMEs including their arrival time at Earth. A number of such studies, based mainly on HI observations, have been performed previously. For example, \citet{Davis2010} applied the Fixed-Phi Fitting method \citep{Sheeley2008.apjl,Rouillard2008,Davies2009} to HI observations to estimate the propagation direction and speed of CMEs and compared their results with those obtained by \citet{Thernisien2009} using Forward Modeling method for the same CMEs observed in the COR FOV. The authors found that their retrieved CME directions were in good agreement with the those from forward modeling, while the discrepancy in speed between the two techniques could be explained in terms of the acceleration of slow CMEs and the deceleration of fast CMEs in the HI FOV. It is also worth noting that \citet{Thernisien2009} had compared their estimated CME propagation directions  with  those  obtained  by \citet{Colaninno2009} using an entirely different technique. The method of \citet{Colaninno2009} is based on the constraint that both COR2-A and COR2-B should record the same true mass for any given CME. \citet{Wood2009} applied the Point-P \citep{Howard2006} and Fixed-Phi methods to the CME of 2008 February 4; their results indicated that the Fixed-phi method performed better, which they used as an argument for  the studied CME having a small angular extent. Similarly, \citet{Wood2010} implemented the Point-P, Fixed-Phi, and Harmonic Mean \citep{Lugaz2009} methods on the 2008 June 1 CME, and showed that different methods can give significantly different kinematic profiles especially in the HI2 FOV. Recently, \citet{Lugaz2010} has assessed the accuracy and limitations of using two fitting methods (Fixed-Phi Fitting and Harmonic Mean Fitting) and two stereoscopic methods Geometric Triangulation (by \citealt{Liu2010}) and Tangent to A Sphere (by \citealt{Lugaz2010.apj}), to estimate the propagation direction of 12 CMEs launched during 2008 and 2009. Their results showed that the Fixed-Phi Fitting approach can result in significant errors in CME direction when the CME is propagating outside 60$\arcdeg$ $\pm$ 20$\arcdeg$ of the Sun-spacecraft direction and Geometric Triangulation can yield large errors if the CME is propagating outside $\pm$ 20$\arcdeg$ of the Sun-Earth line. More recently, \citet{Colaninno2013} derived the deprojected height-time profiles of CMEs by applying the graduated cylindrical shell model (Forward Modeling: \citealt{Thernisien2009}) using SECCHI and LASCO images. In their study, they fitted the derived height-time data with six different methods to estimate the CME arrival time and speed at Earth.

The aforementioned studies are mainly limited to making comparisons of CME propagation directions, and sometimes speeds, retrieved using different methods. These studies do not consider the characteristics of either the individual CMEs or the ambient medium into which they  are launched, nor do they (except \citealt{Colaninno2013}) use the estimated kinematics of CMEs to predict their arrival time near 1 AU. Neither do they compare the derived CME arrival time with the actual CME arrival time identified from in situ observations.  However to understand the validity of these techniques for space weather forecasting, which has significant consequences for life and technology in space and on Earth, one should use them to predict the arrival time  and speed of CMEs launched at different speeds, launched  into different ambient solar wind conditions. Our selection of three CMEs, launched with different speeds into different solar wind conditions (on 2010 October 6, April 3 and February  12), satisfy such criteria. Previously studies have tended to apply only a limited number of techniques to any individual CME. Here we implement a total of 10 techniques, ranging from single spacecraft methods and their fitting analogues to stereoscopic techniques, to the CMEs under study; such extensive analysis as is undertaken in our study has not previously been reported.  Moreover, we not only compare the estimated direction and speed of the three selected CMEs, but also assess the relative performance of these techniques in estimating CME arrival time at L1. We compare the estimated arrival time and speed of each CME to the actual arrival time and speed based on in situ signatures at L1 \citep{Zurbuchen2006}. Such a study is a useful step towards identifying the most appropriate techniques for the practical purpose of forecasting CME arrival time at Earth, in the near future.

\section{Methodology and its application to selected CMEs}

The main objective of the STEREO mission is to improve our understanding of the initiation  and evolution of CMEs, with particular emphasis on Earth-directed events. After its launch, various reconstruction techniques were developed to estimate the 3D kinematic properties of CMEs near the Sun using SECCHI/COR1 and COR2 observations, and further from the Sun using SECCHI/HI1 and HI2 observations. Note that, even prior to the launch of STEREO, some CMEs had been tracked out to large distances from the Sun using SMEI observations \citep{Tappin2004,Jackson2006,Howard2006,Webb2006,Kahler2007}. In the work reported in the current paper, we applied ten techniques to three CMEs observed by STEREO on 2010 October 6, 2010 April 3 and 2010 February 12. The CME of 2010 October 6 was a slow speed CME, while the other two were observed to be fast in the COR2 FOV. The CMEs of October 6 and February 12, traversed through a slow speed ambient solar wind medium. The fast 2010 April 3 CME experienced only a modest deceleration as it propagated through the IP medium. It was the fastest CME at 1 AU since the 2006 December 13 CME. We note that, during this time, the Earth was in the throes of high speed solar wind which perhaps governed the dynamics of this CME. The 2010 April 3 CME has been investigated extensively in earlier studies \citep{Mostl2010,Liu2011,Wood2011,Temmer2011,Mishra2013}. 

The methods used to derive the heliospheric kinematics of CMEs  can be classified into two groups, one which uses single spacecraft observations and the other which requires simultaneous observations from two viewpoints. We have implemented seven single spacecraft methods,  Point-P (PP: \citealt{Howard2006}), Fixed-Phi (FP: \citealt{Kahler2007}), Harmonic Mean (HM: \citealt{Lugaz2009}), Self-similar expansion (SSE: \citealt{Davies2012}), Fixed-Phi Fitting (FPF: \citealt{Rouillard2008}), Harmonic Mean Fitting (HMF: \citealt{Lugaz2010}) and Self-Similar Expansion Fitting (SSEF: \citealt{Davies2012}), which are based on data from a single viewpoint. We have also implemented three stereoscopic methods, namely, Geometric Triangulation (GT: \citealt{Liu2010}), Tangent to A Sphere (TAS: \citealt{Lugaz2010.apj}) and Stereoscopic Self-Similar Expansion (SSSE: \citealt{Davies2013}), which require simultaneous observations from two viewpoints. Using the FP, HM and SSE methods, one can estimate the kinematics of a CME, provided an estimate of its 3D propagation direction is known. The  3D propagation direction of a CME close to the Sun can be estimated by applying the scc$\_$measure.pro routine to COR2 observations; this method is based on an epipolar geometry \citep{Inhester2006} and is available in the solar-soft library for 3D reconstruction \citep{Thompson2009}.  We derive the kinematics of these three CMEs in the heliosphere using the aforementioned methods and use those kinematics (averaged over the last few data points) as inputs to the Drag Based Model (DBM: \citealt{Vrsnak2013}) to estimate the arrival times of the CMEs at L1. If, for any method, the derived kinematics show implausible (unphysical) variations over the last few points then the kinematics prior to that time are used instead.

When a CME is far from the Sun, the Lorentz and gravity forces decrease such that drag can be considered to govern CME dynamics. Although it is not proven that drag is the only force that shapes  CME dynamics in the IP medium, the observed deceleration/acceleration of some CMEs has been closely reproduced by considering only the drag force acting  between the CME and the  ambient  solar  wind medium \citep{Lindsay1999, Cargill2004, Manoharan2006, Vrsnak2009, Lara2009}. In our study, we have used the DBM to derive the kinematic properties for the distance range beyond which a CME cannot be tracked in the J-maps. The CMEs under study have been tracked to sufficiently large distances that, beyond these distances, we consider it appropriate to assume that only drag forces act on the CMEs. Therefore we use the drag model for estimating the  arrival times of the CMEs at L1. The DBM model is based on the assumption that, after 20 R$_\sun$, the dynamics of CMEs is solely governed by the drag force and that the drag acceleration has the form, $a_{d}$ = -$\gamma$ $(v-w)$ $|(v-w)|$, (see e.g. \citealt{Cargill1996, Cargill2004,Vrsnak2010}), where $v$ is the speed of the CME, $w$ is the ambient solar wind speed and $\gamma$ is the drag parameter. The drag parameter is given by $\gamma$ = $\frac{c_{d}A \rho_{w}}{M + M_{v}}$, where c$_{d}$ is the dimensional drag coefficient, $A$ is the cross-sectional area of the CME perpendicular to its propagation direction (which depends on the CME-cone angular width), $\rho_{w}$ is the ambient solar wind density, $M$  is the CME mass, and $M_{v}$  is the virtual CME mass. The latter is written as, $M_{v}$ = $\rho_{w} V/2$, where $V$ is the CME volume. The statistical study of \citet{Vrsnak2013} showed that the drag parameter generally lies between 0.2 $\times$ 10$^{-7}$ and 2.0 $\times$ 10$^{-7}$ km$^{-1}$. They assumed that the mass and angular width of CMEs do not vary beyond 20 R$_\sun$ and also showed that the solar wind speed should be selected to lie between 300 and 400 km s$^{-1}$ for slow solar wind conditions. \citet{Vrsnak2013} also showed that the ambient solar wind speed should be chosen to lie between 500 and 600 km s$^{-1}$, along with a lower value of the drag parameter, for the case where the CME propagates in a high speed solar wind or if a coronal hole is present in the vicinity of the CME source region. Although a given CME will be associated with a drag parameter of a particular value (based on its mass and cone angular width), we consider the entire statistical range of the drag parameter derived by \citet{Vrsnak2013} when applying the DBM to the three CMEs in our study. This is done due to the lack of certainty of the estimated CME characteristics (e.g. cone angle, mass) on which the drag parameter depends \citep{Vourlidas2000, Colaninno2009}.

In the following sections, we briefly revisit various reconstruction techniques and discuss their application to the three selected CMEs. The analysis of  the CME of 2010 October 6 is presented in Section 2.1 in detail. Results of analogous analyses of the 2010 April 3 and February 12 CMEs are presented briefly in Sections 2.2 and 2.3, respectively.

\subsection{2010 October 6 CME}

\begin{figure}
\begin{center}
\includegraphics[angle=0,scale=.40]{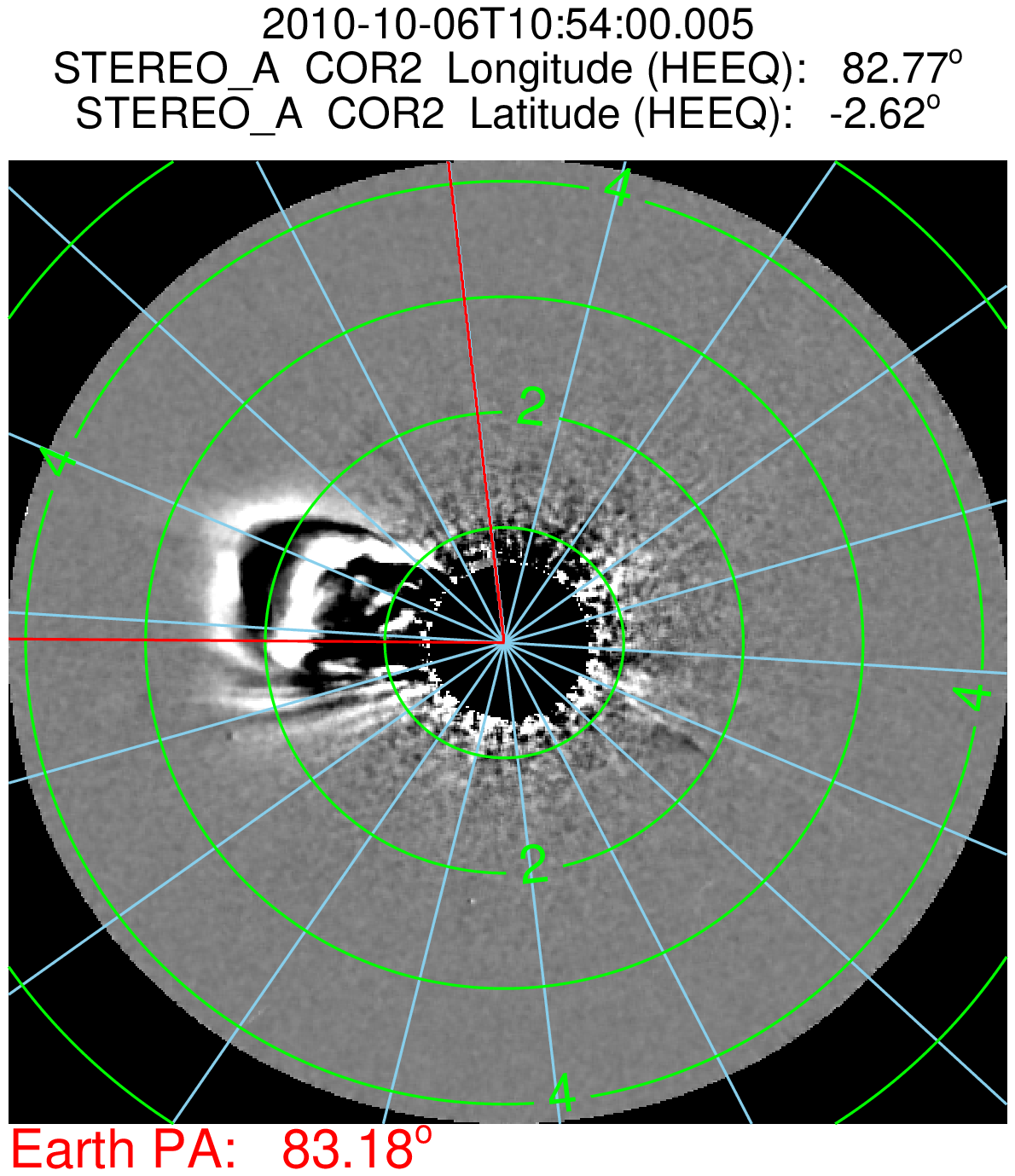}
\includegraphics[angle=0,scale=.40]{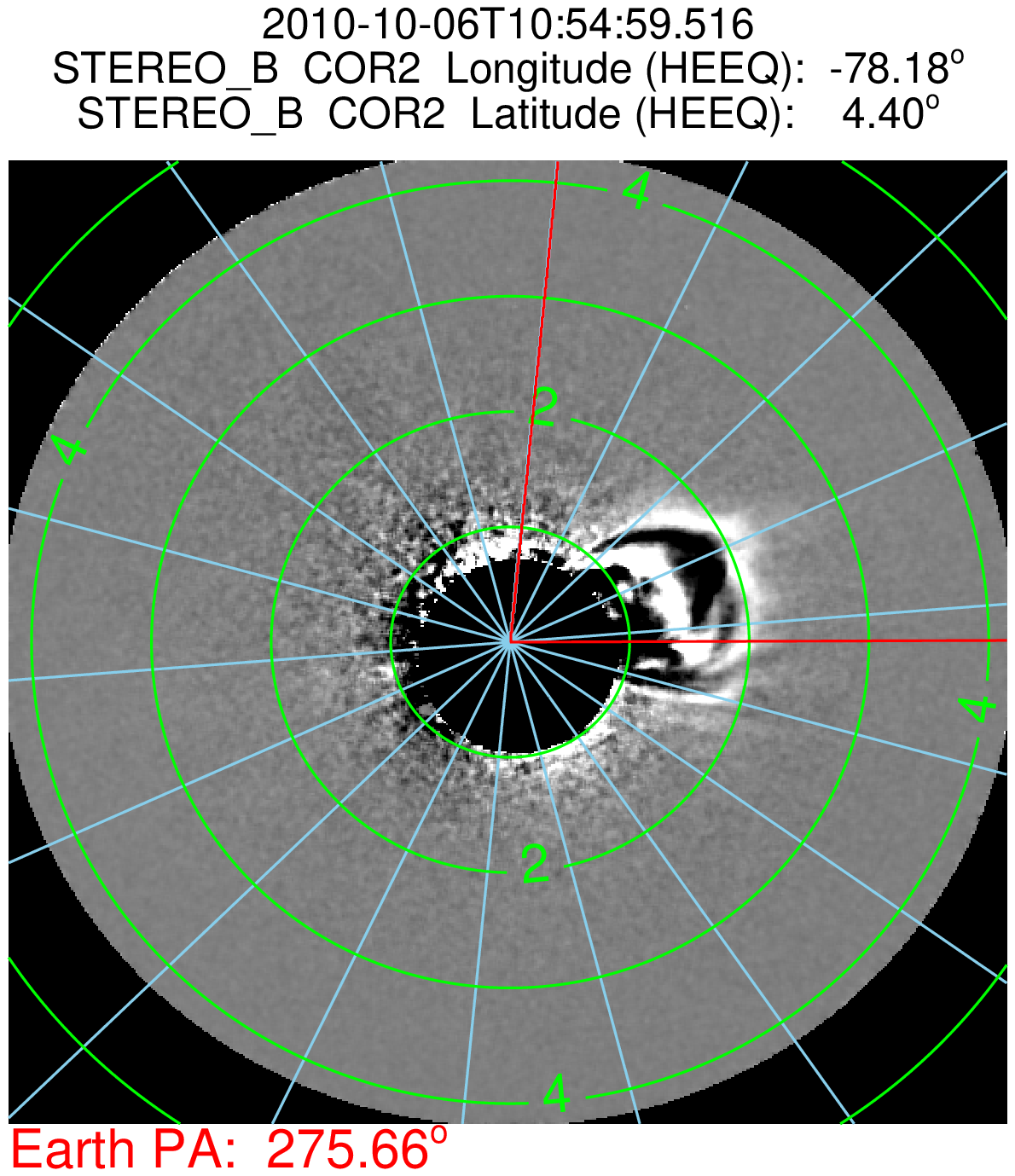}  \\
\includegraphics[angle=0,scale=.40]{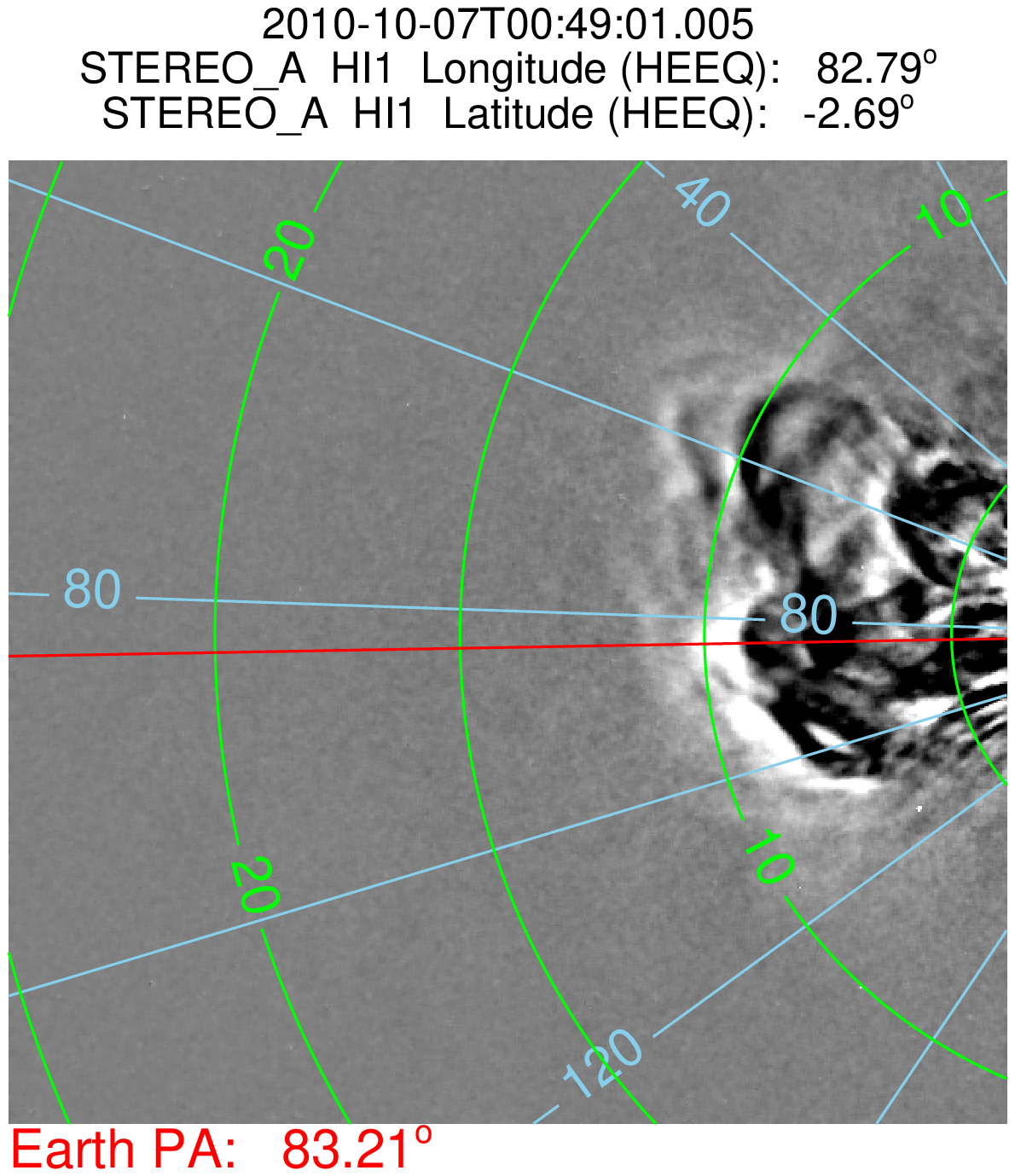}
\includegraphics[angle=0,scale=.40]{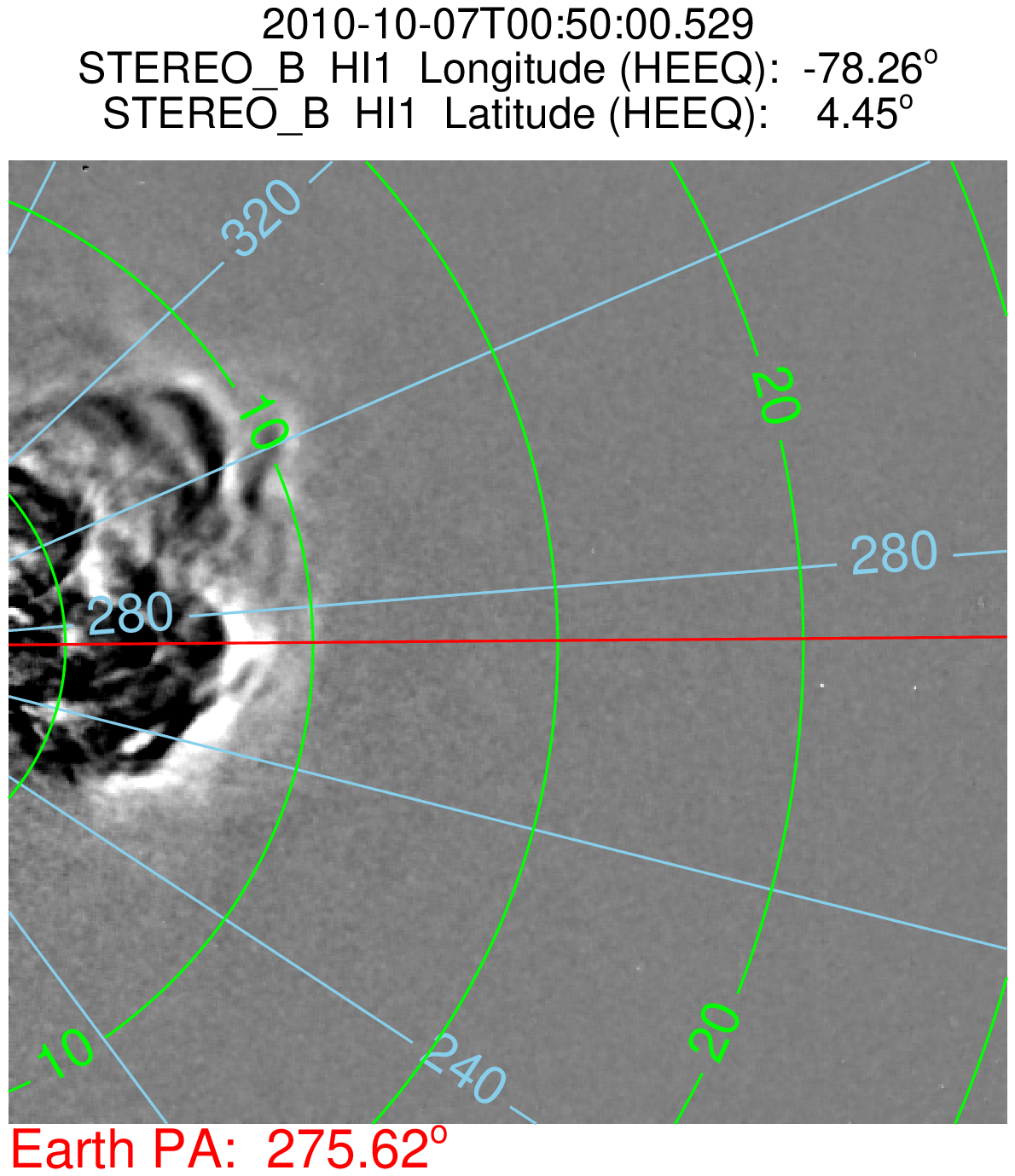}  \\
\includegraphics[angle=0,scale=.40]{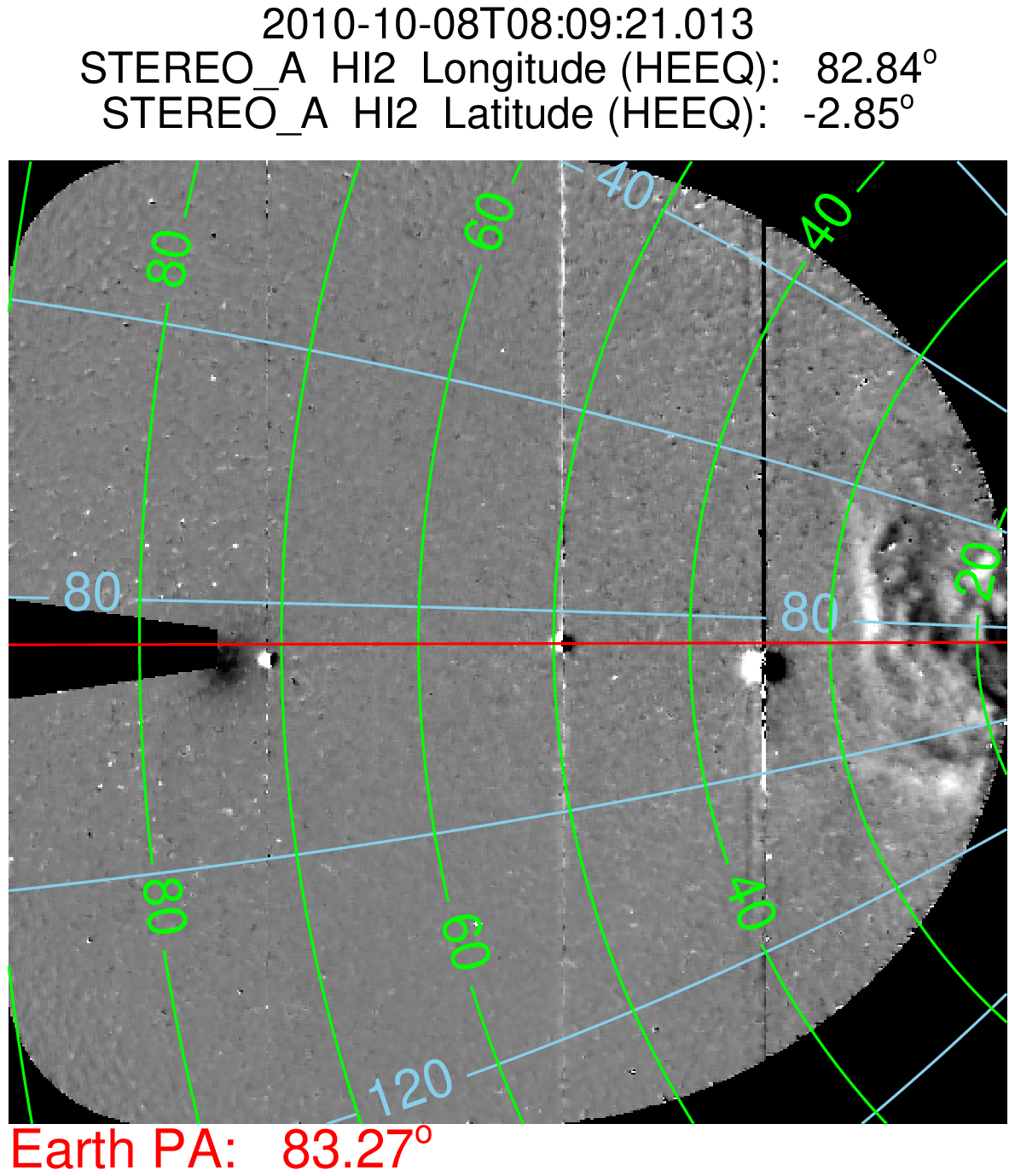}
\includegraphics[angle=0,scale=.40]{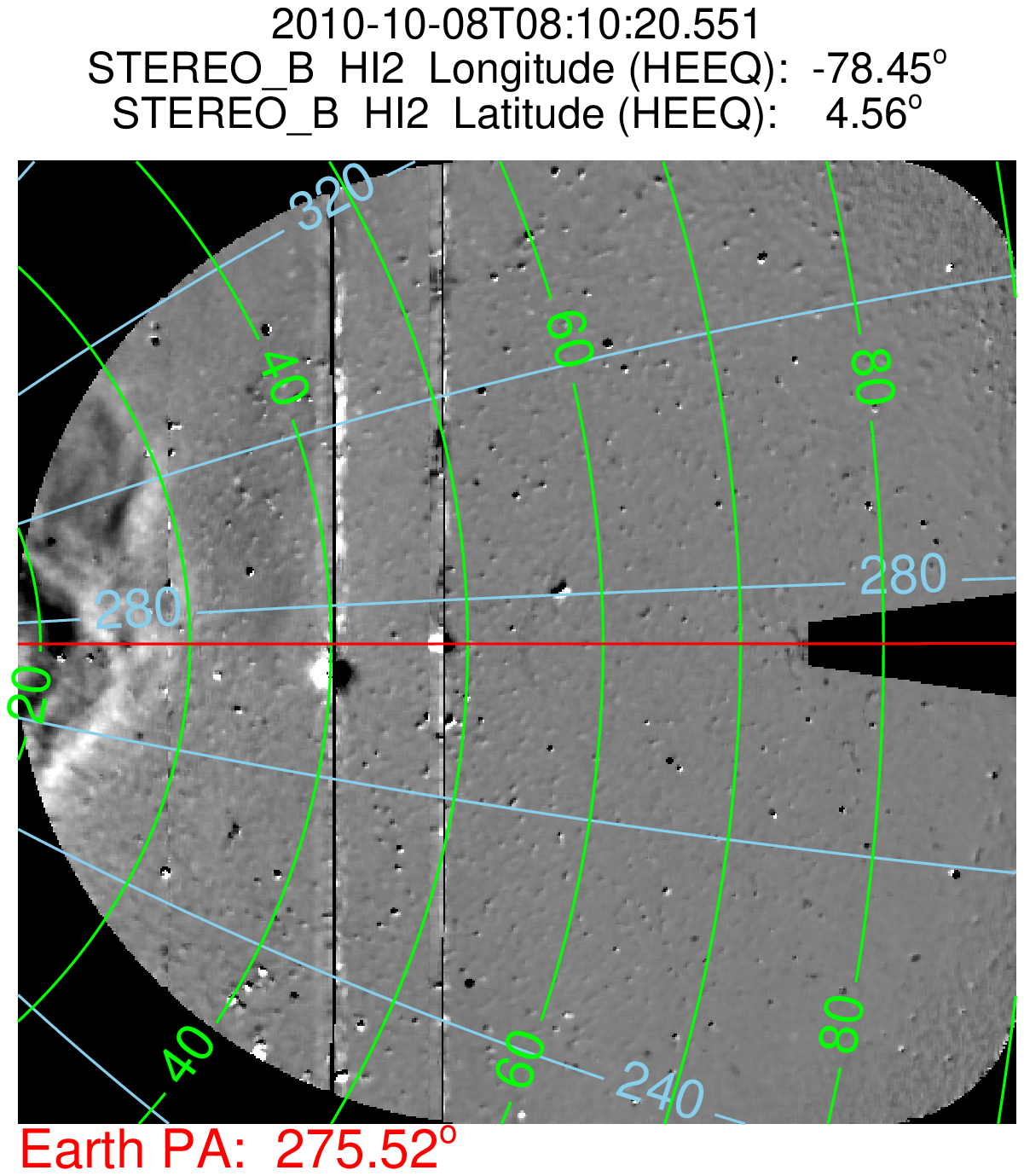}
\caption{\scriptsize{Evolution of the 2010 October 6 CME observed in COR2, HI1 and HI2 images from STEREO-A (left column) and B (right column). Contours  of elongation angle (green) and position angle (blue) are overplotted. The vertical red line in the COR2 images marks the 0$\arcdeg$ position angle contour. The horizontal lines (red) on all panels indicate the position angle of Earth.}}
\label{Evolution}
\end{center}
\end{figure}

The CME of 2010 October  6 was first observed by SOHO/LASCO C2 at 07:12 UT as a partial halo with a linear (POS) speed of 282 km s$^{-1}$ (online LASCO CME catalog: http://cdaw.gsfc.nasa.gov/CME$\_$list/; see \citealt{Yashiro2004}). This CME was associated with a filament eruption from the north-east (NE) quadrant of the solar disc. In the LASCO C3 FOV, this CME was observed to accelerate at 7 m s$^{-2}$. SECCHI/COR1-A and COR1-B, with an angular separation of approximately 161$\arcdeg$, first observed the CME at 04:05 UT in the NE and north-west (NW) quadrants, respectively. The CME was subsequently observed by COR2, and HI1 and HI2, on both STEREO-A and B (Figure~\ref{Evolution}). The arrival of this CME at the Earth caused a moderate geomagnetic storm with a peak Disturbance Storm Time (Dst) index of approximately -80 nT on 2010 October  11 at 19:00 UT. 

We applied the COR2 data processing scheme as described by \citet{Mierla2010} before implementing  the tie-pointing technique of 3D reconstruction \citep{Thompson2009}. The 3D radial speed of the 2010 October 6 CME is estimated to be 340 km s$^{-1}$  at a 3D height of nearly 10 R$_\sun$ from the Sun. The central latitude of the CME feature was estimated to be $\approx$ 20$\arcdeg$ North and its longitude  $\approx$ 10$\arcdeg$ East of the Earth. As the CME was propagating only slightly north-eastward of the Sun-Earth line, it was likely to impact the Earth (which it did).  Assuming a constant speed of 340 km s$^{-1}$ beyond the COR2 FOV, the predicted CME arrival time at L1 is estimated to be on 2010 October 11 at 06:10 UT. 

\subsubsection{Reconstruction techniques using single spacecraft observations}
Using heliospheric imagers, which image at and across large distances from the Sun, 3D information about CMEs can be extracted without the need for multiple viewpoint observations. This is possible because of the fact that, when CMEs are far from the Sun, the geometrical and Thomson scattering linearity that is often assumed near the Sun break down \citep{Howard2011}. In this section, we apply reconstruction methods based on single viewpoint observations of CMEs in the heliosphere, i.e.  the PP, FP, HM and SSE methods mentioned in Section 2.

\subsubsubsection{Point-P (PP) method using SECCHI/HI observations}

Soon after the launch of SMEI \citep{Eyles2003}, which can provide the elongation angle (Sun-observer- feature angle) of a solar wind feature such as a CME, a standard technique based on known assumptions was applied to convert the elongation angle to a distance from Sun center; this conversion methodology was termed the Point-P (PP) method \citep{Howard2006,Howard2007}. The accuracy of this conversion is constrained by the effects of the Thomson scattering process and the geometry of CMEs, which govern their projection in the images. In the PP method, it is assumed that a CME forms a circular structure centered on the Sun and an observer tracks the point where the CME intersects the Thomson surface \citep{Vourlidas2006}. Under these assumptions, the derived radial distance of the CME from the Sun center is  R$_{PP}$ = $d$ Sin$\alpha$, where $\alpha$ is the measured elongation of the CME and $d$ is the distance of the observer from the Sun. In the case where the small (elongation) angle approximation can be applied, the PP method is close to the POS approximation. Since the earliest days of CME observations, the POS approximation has been used to convert coronagraph measurements of a CME feature to physical distance from the Sun.   

However, \citet{Howard2012} and \citet{Howard2013} have de-emphasized this concept by showing that the maximum intensity of scattered light per unit density is spread over a broad range of scattering angles (called Thomson plateau).  They conclude that CME features can be observed far from the Thomson surface and that their detectability is governed by the location of the feature relative to the plateau rather than the Thomson surface. The existence of this Thomson plateau and the oversimplified CME geometry assumed in the PP method are likely to lead to significant errors in estimated kinematics.    

\begin{figure}
\begin{center}
\includegraphics[angle=0,scale=.50]{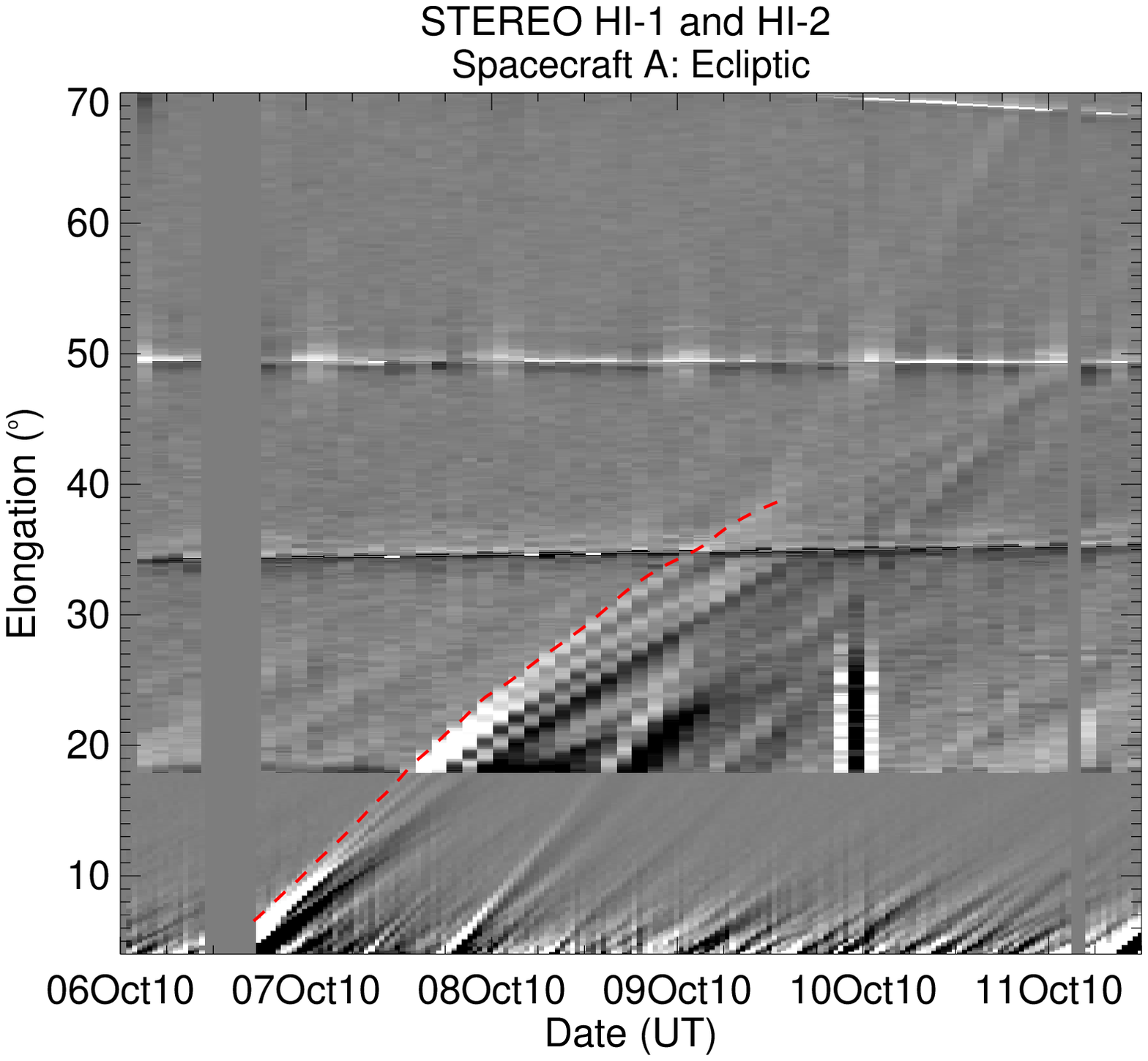}
\includegraphics[angle=0,scale=.50]{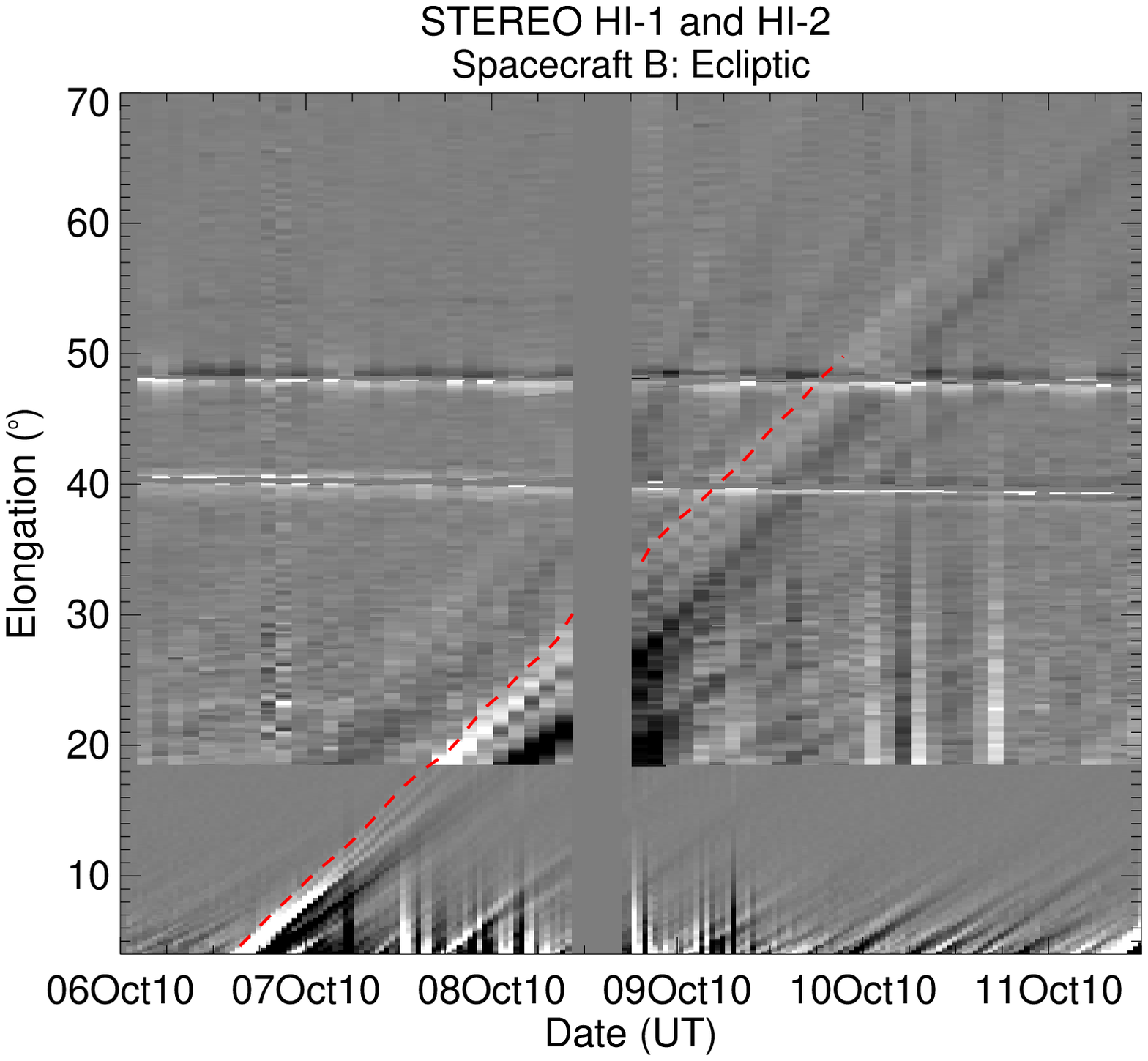}
\caption{\scriptsize{Ecliptic time-elongation maps (J-maps) for STEREO-A (left) and STEREO-B (right) constructed from running differences images from HI1 and HI2, for the time interval extending from 06 Oct 2010 to 11 Oct 12:00 UT, 2010. The leading edge of the bright feature (corresponding to the leading edge of the initial CME front) is tracked in the J-maps (red lines)}}
\label{Jmaps}
\end{center}
\end{figure}

When a CME is in the FOV of a heliospheric imager (Coriolis/SMEI or SECCHI/HI), it becomes very diffuse. Therefore to track and extract the time-elongation profile of a moving solar wind structure, a technique developed originally by \citet{Sheeley1999}, involving the generation of time-elongation maps (J-maps: \citealt{Rouillard2008, Davies2009, Mostl2010, Liu2010, Harrison2012}) is often applied. For the current study, we constructed J-maps using running difference images from the HI1 and HI2 instruments, as explained in \citet{Mishra2013}. Ecliptic J-maps covering the passage of the 2010 October 6 CME, from the viewpoints of both STEREO-A and  B, are shown in Figure~\ref{Jmaps}. In the HI2-A FOV on 2010 October 6, Venus, Earth, Jupiter, and Neptune were identified at position angles of 86.5$\arcdeg$, 83.1$\arcdeg$, 84.4$\arcdeg$, and 83.8$\arcdeg$ with elongations of 34.3$\arcdeg$, 49.5$\arcdeg$, 73.1$\arcdeg$, and 50.3$\arcdeg$ respectively (see Figure~\ref{Evolution}). In the HI2-B FOV, Venus and Earth were identified at elongations of 40.4$\arcdeg$, and 48.2$\arcdeg$ respectively. In each J-map (Figure~\ref{Jmaps}), there are two horizontal lines that are due to presence of Venus and Earth in the HI2 FOV. A slanted line that appears in the STEREO-A J-map on October 10 is due to entrance of Jupiter into the HI2-A FOV.

\begin{figure}
\begin{center}
\includegraphics[angle=0,scale=.50]{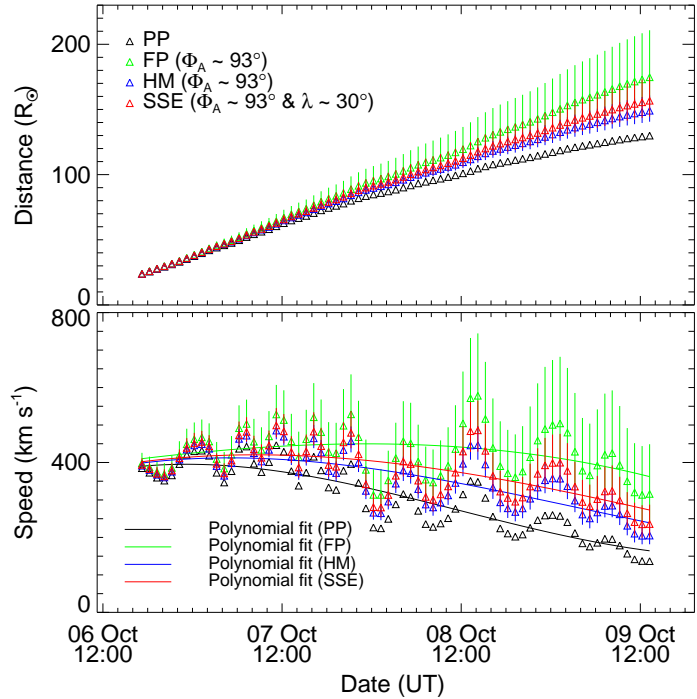}
\caption{\scriptsize{The derived distance profiles based on application of the PP, FP, HM and SSE methods for the tracked feature are shown in the top panel. The bottom panel presents speed profiles derived from the adjacent distances using three point Lagrange interpolation (solid  line shows the polynomial fit). Vertical lines show the errors bars, calculated using  propagation  directions that are +10$\arcdeg$ and -10$\arcdeg$ different to  the  value ($\phi$) estimated  using  tie-pointing, as described in Section 2.2.1.}}
\label{STAA06Oct}
\end{center}
\end{figure}

In each J-map, a set of the positively inclined bright features corresponds to the 2010 October 6 CME. We tracked the leading edge of first bright CME feature, corresponding to the initial CME front, in the STEREO-A J-map (red dashed line in the left panel in Figure~\ref{Jmaps}). We manually extracted the time-elongation profile of this outward-moving feature and applied the PP approximation to the elongation data based on the expression quoted earlier in this section. The CME front can be tracked out to 39$\arcdeg$ elongation in the STEREO-A J-map. The derived radial distance and speed of the tracked feature are plotted in Figure~\ref{STAA06Oct} in black. It is to be noted that, using this method, we obtain time variations of CME radial distance and speed; this is not possible using the single spacecraft fitting techniques (FPF, HMF and SSEF) which give only a single value of the radial speed unless we apply the triangulation approach. Note, however, that an estimation of the CME propagation direction is not retrievable using the PP approach. The speed is calculated from adjacent estimates of distance using the IDL $deriv$ function, which performs a three point Lagrange interpolation on the data points to be differentiated. The estimated speed variation in Figure~\ref{STAA06Oct} suggests that the tracked feature undergoes  deceleration.

We input the CME kinematics, estimated by implementing the PP method, into the DBM to predict the arrival time of the CME at L1. The CME front is tracked out to a heliocentric distance of 129.9 R$_\sun$ (0.6 AU) on 2010 October 9 at 09:14 UT, where it has a radial speed of 180 km s$^{-1}$.  The kinematics at the furthest distances to which this feature can be tracked are used as inputs to the DBM, along with the minimum and maximum values of the drag parameter estimated  by \citet{Vrsnak2013} and an ambient solar wind speed of 350 km s$^{-1}$. The predicted arrival time and transit speed of the tracked CME front at L1, corresponding to the extreme values of the drag parameter, are given in Table 1.

From Figure~\ref{Jmaps}, we notice a data gap in the STEREO HI-B observations and that the J-map quality is slightly better for STEREO HI-A. An important reason for the poorer quality of the HI-B J-maps is that the HI2-B images are out of focus compared to HI2-A images \citep{Brown2009}. Another reason for the poorer quality of HI-B images is that HI on STEREO-B suffers small pointing discontinuities due to dust impact. Since HI-B is facing the direction of travel of the STEREO-B spacecraft, it gets impacted directly by interplanetary dust; HI-A is on the opposite side to the direction of travel of STEREO-A, so does not suffer direct impact \citep{Davis2012}. Due to the large gradient inherent in the F-coronal signal, even a small pointing offset can result in an inaccurate F-corona subtraction, which results in degraded image quality. Despite this, we tracked the leading edge of the October 6 CME in HI-B J-maps even beyond the data gap (right panel in Figure~\ref{Jmaps}). Adopting the same procedure described for the feature tracked by STEREO-A, we also estimated the kinematics and the arrival times at L1 of the CME based on its elongation profile extracted from the STEREO-B J-map. Its estimated kinematics over the last few tracked points, which are used as inputs to the DBM, and predicted arrival time at L1 are noted in Table 1.

\subsubsubsection{Fixed-Phi (FP) Method using SECCHI/HI observations}
After the advent of truly wide-angle imaging, \citet{Kahler2007} developed a different method to convert elongation to radial distance, by assuming that the solar wind feature (such as a CME) can be considered as a point source moving radially outward in a fixed direction ($\phi_{FP}$) relative to an observer located at a distance $d$ from the Sun.

The distance profile of the CME can be derived using this so called Fixed-Phi (FP) approximation by assuming a propagation direction that can be determined by identifying the CME's source region. However, in our analysis of the CME of 2010 October 6, we actually use the  propagation direction ($\phi_{FP}$) derived from the 3D reconstruction of COR2 data as discussed in Section 2.1. We assume that, beyond the COR2 FOV, the CME will continue to travel in the same direction. The estimated longitude of the CME is $\approx$ 10$\arcdeg$ East of the Earth, which corresponds to a longitude difference of $\approx$ 93 $\arcdeg$ from the STEREO-A spacecraft; the separation angle between STEREO-A and the Earth was $\approx$ 83$\arcdeg$ at that time. Using the elongation variation of the tracked feature, extracted manually from the ecliptic J-maps constructed from HI-A images as shown in Figure~\ref{Jmaps} (left), and $\phi_{FP}$ = 93$\arcdeg$, we calculated the distance profile of the leading bright front of the CME. The obtained time variations of the radial distance and speed are shown in green in Figure~\ref{STAA06Oct}. The unphysical deceleration of the CME beyond 100 R$_{\sun}$ (suggested by a speed less than ambient solar wind speed) may be due to the erroneous fixing of the propagation angle in the case of a real deflection or, indeed, the inaccurate characterization of the propagation angle. However, it is most likely due to breakdown of the simple assumption that observer is always looking at the same point-like feature of the CME. This can lead to large errors in the estimated height of the CME leading edge, particularly at greater elongations where the expanding CME geometry plays a significant role and the observer will be unlikely to record the intensity from the same part of CME's leading edge (artificial deflection: \citealp{Howard2009, Howard2011}). The limitations of this and other methods will be discussed in Section 4.

As with the PP analysis, we apply the DBM based on the kinematics of the tracked CME feature estimated using the FP method. Running the DBM with the derived CME kinematics as inputs, along with an ambient solar wind speed of 350 km s$^{-1}$ and the two extreme values of the drag parameter, we obtained the L1 arrival times and transit speeds given in Table 1.

We also applied the FP method to the elongation profile derived from the STEREO-B J-map (Figure~\ref{Jmaps}, right). From STEREO B, the feature can be tracked out to 162 R$_\sun$ (0.75 AU) on 09 October 07:27 UT, where its speed is approximately 435 km s$^{-1}$.1 Again, these kinematics are used as inputs to the DBM to estimate the arrival times and transit speeds at L1 (see Table 1).

\subsubsubsection{Harmonic Mean (HM) Method using the SECCHI/HI observations}
To convert elongation angle to radial distance from the center of the Sun, \citet{Lugaz2009} assumed that a CME can be represented as a self - similarly expanding sphere attached to Sun-center, with its apex traveling in a fixed radial direction. The authors further assumed that an observer measures the scattered emission from that portion of the sphere that the line-of-sight intersects tangentially. Based on these assumptions, they derived the distance ($R_{HM}$) of the apex of the CME from Sun-centre as a function of elongation and found that this distance is the harmonic mean of the distances estimated using the FP and PP methods. Hence the method is referred to as the Harmonic Mean (HM) method \citep{Lugaz2009}.

We used the CME longitude estimated from 3D reconstruction of COR2 data (as described earlier), and the elongation profile extracted from the STEREO-A J-map, to estimate the distance and speed profiles of the CME front using the HM approximation (blue lines in Figure~\ref{STAA06Oct}). A polynomial fit to the speed profile (solid blue line in bottom panel of Figure~\ref{STAA06Oct}), suggests an overall deceleration of the tracked CME feature. The non-physical deceleration of this feature at large distances could possibly result from the real deflection of this feature or an inaccuracy in the assumed `fixed' direction, or be due to the fact that the observer (in this case STEREO-A) detects scattered light from a different part of the CME than that assumed (artificial deflection). The  limitations of the Harmonic mean method are discussed in Section 4.

Again, the estimated kinematics derived over the last segment of the tracked time-elongation profile at around 13:14 UT on 2010 October 9 (distance: 149 R$_\sun$, 0.69 AU and speed 230 km s$^{-1}$) are used as inputs into the DBM, to predict the CME arrival time and transit speed at L1 (Table 1). As above, the DBM assumes an ambient solar wind speed of 350 km s$^{-1}$, and is run for the two extreme values of the drag parameter.

The same methodology is applied to the CME track observed by STEREO-B. At the end of its observed track, on October 9 at 21:27 UT, the CME's speed is estimated to be 410 km s$^{-1}$ at a distance of 189 R$_\sun$ (0.87 AU). These values are used as inputs in the DBM to derive arrival times and transit speeds at L1 (Table 1).

\subsubsubsection{Self-Similar Expansion (SSE) method using SECCHI/HI observations}
Motivated by the \citet{Lugaz2010.apj} ``model 2'' geometry, as an expanding circle not anchored to the Sun, \citet{Davies2012} derived an expression for the elongation variation as a function of time of a CME conforming to such a geometry viewed from a single vantage point (equation 6 of \citealt{Davies2012}). \citet{Davies2012} termed this model the Self-Similar Expansion (SSE) model. In its extreme forms, the SSE geometry - which is characterized by an angular half-width $\lambda$ - is equivalent to the FP ($\lambda$ = 0$\arcdeg$) and HM models ($\lambda$ = 90$\arcdeg$). Assuming $\lambda$ = 30$\arcdeg$, we use the longitude estimated from 3D reconstruction in the COR2 FOV to estimate the distance and speed profiles of the CME tracked in the STEREO-A J-map, using the SSE method (shown in red in Figure~\ref{STAA06Oct}). The DBM was run in the same manner as described earlier, based on these kinematics inputs, to obtain the arrival times and transit speeds (Table 1). The same methodology was applied to STEREO-B observations (results are included in Table 1).

\subsubsubsection{Error analysis for FP, HM and SSE method}

As described above, we used, as input to the FP, HM and SSE method, the propagation direction ($\phi$) of the CME  estimated from the  tie-pointing  method  of 3D  reconstruction. To examine the uncertainties arising from the use of the tie-pointing method, we compared the CME kinematics derived using propagation directions obtained from other methods.  We note that, for all CMEs that form part of this study, the propagation directions estimated using tie-pointing  and forward modeling \citep{Thernisien2009} are  within  10$\arcdeg$, which  is  in agreement  with  results of  \citet{Mierla2010}. Propagation directions estimated from CME source location identification are also within 10$\arcdeg$ of the values obtained from tie-pointing.  We repeated our FP, HM and SSE analysis, as described  above,  using  propagation  directions  that are +10$\arcdeg$ and -10$\arcdeg$ different to  the  value ($\phi$) estimated  using  tie-pointing in order to estimate uncertainties in distance (vertical error bars in Figure~\ref{STAA06Oct}). The standard deviation (uncertainty) in the speed is calculated using the IDL $derivsig$ function.  These error bars do not denote all errors in these single spacecraft methods, but simply represent the sensitivity of the method to uncertainties in direction.

\subsubsubsection{Fixed-Phi Fitting (FPF), Harmonic Mean Fitting (HMF) and Self-Similar Expansion Fitting(SSEF) method using SECCHI/HI observations}
The original concept of \citet{Sheeley1999}, exploiting the deceptive acceleration or deceleration in the time-elongation profiles of CMEs moving in a fixed direction with a constant speed out to large elongation angles, is used widely to estimate the direction and speed of CMEs. This initial concept forms the basis of several single spacecraft fitting techniques, based on the aforementioned assumed CME geometries \citep{Rouillard2008, Sheeley2008, Davis2009, Mostl2009, Mostl2010, Howard2009, Mostl2011}. In Fixed-Phi Fitting (FPF), the theoretical elongation variation, characterized by equation (1) of \citet{Rouillard2008}, which fits best the observed elongation variation provides values for the most physically realistic combinations of speed ($v_{FP}$), direction ($\phi_{FP}$), and launch time from Sun-centre ($t_{0FP}$), where $\alpha(t_{0FP})$ = 0. FPF has been applied to CIRs (e.g. \citealt{Rouillard2008}) and CMEs (e.g. \citealt{Davis2009, Davies2009, Rouillard2009}).

\begin{figure}
\begin{center}
\includegraphics[angle=0,scale=.50]{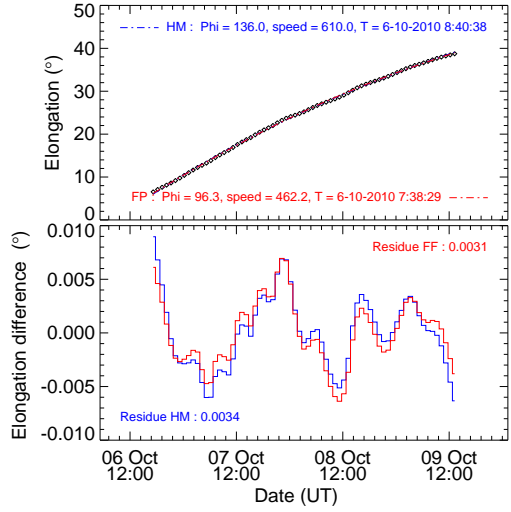} 
\caption{\scriptsize{Best-fit FPF  and HMF  results are shown with red and blue colors, respectively, for the tracked CME feature. In the top panel, best-fit theoretically obtained elongation variations are shown. In the bottom panel, residuals between the best-fit theoretical elongation variations and the observed elongation variations are shown.}}

\label{FF_HMF}
\end{center}
\end{figure}

We applied the FPF technique to the elongation variation of the October 6 CME derived from the STEREO-A HI J-map. We implemented the IDL routine MPFITFUN \citep{Markwardt2009} to find the set of $v_{FP}$, $\phi_{FP}$ and $t_{0FP}$ parameters that best reproduced the observed elongation variation. In the upper panel of Figure~\ref{FF_HMF} (red line), we show how well the observed variation is reproduced by equation (1) of \citet{Rouillard2008}. We estimated the fitting residuals (bottom panels) by following the approach of \citet{Mostl2011}. Applying FPF to the initial front of the October 6 CME tracked in the STEREO-A J-map yields a propagation direction of 96.3$\arcdeg$ from the spacecraft (i.e. 13.3$\arcdeg$ East of the Sun-Earth line), a speed of 462 km s$^{-1}$, and a launch time of 07:38 UT on 2010 October 6. Assuming this speed as constant from the Sun to L1, the CME is predicted to arrive at L1 at 00:39 UT on 2010 October 10.

\citet{Lugaz2010} derived an expression for the elongation variation, with time, based on the HM geometrical model (their equation 2).  This equation forms the basis of the Harmonic Mean Fitting (HMF) technique. Subsequently, \citet{Mostl2011} derived a different form for the HMF relation (their equations 4 and 5), which we have used to implement the HMF technique. Following the same basic procedure as for FPF, the best fit elongation variation, derived using the HMF approach, is shown in Figure~\ref{FF_HMF} (upper panel) in blue. Applying the HMF technique to the time-elongation profile of the CME tracked in STEREO-A gives a propagation direction of 136$\arcdeg$ from the spacecraft, i.e. 53$\arcdeg$  East of the Sun-Earth line, a speed of 610 km s$^{-1}$, and a launch time of 08:40 UT on 2010 October 6. This speed (when corrected for off-axis propagation, see below) gives a predicted L1 arrival time of 00:19 UT on October 11.

Equation (8) of \citet{Davies2012} can theoretically be used to simultaneously retrieve the best fit launch time, propagation direction, speed and half angular width of a solar wind transient in a procedure termed Self-Similar Expansion Fitting (SSEF). As recommended by those authors, however, we fix the value of half angular width (in our case to 30$\arcdeg$) and applied the SSEF technique to retrieve the best-fit speed, direction and launch time in a similar manner to FPF and HMF. The best fit parameters, and the estimated predicted L1 arrival time, for the October 6 CME based on SSEF are given in Table 1.

The FPF, HMF and SSEF methods are also applied to the elongation variation of the CME extracted from the STEREO-B J-map. The retrieved  best fit parameters (launch time, propagation direction from the observer and speed) and arrival time at L1 are also given in Table 1.

We emphasize that the HMF and SSEF methods estimate the propagation speed of the CME apex and, to estimate its speed in an off-apex direction, a geometrical correction must be applied. The off-apex corrections applicable to the HM and SSE geometries are given by equation (8) of \citet{Mostl2011} and equation (18) of \citet{Mostl2013}, respectively. The speed of the CME in the off-apex direction is less than its speed derived in the apex direction. As the CME apex directions derived from both the HMF and SSEF technique are offset from the Sun-Earth line, we used the geometrically corrected speed to obtain the predicted arrival time of the CME at L1 point. It is this corrected speed that is compared later to the speed measured in situ at L1. Such a geometrical correction is not applicable to the FPF technique, as the CME is assumed to be a point.

\subsubsection{Stereoscopic reconstruction techniques}

A number of stereoscopic techniques have also been developed to determine the distance, speed and direction profiles of CMEs based on simultaneous observations from the two viewpoints of STEREO. In this section, we apply three such methods to determine the kinematics of the 2010 October 6 CME, namely the Geometric Triangulation (GT) technique \citep{Liu2010}, the Tangent to A Sphere (TAS) method \citep{Lugaz2010.apj} and the Stereoscopic Self-Similar Expansion (SSSE) method \citep{Davies2013}.

\begin{figure}
\begin{center}
\includegraphics[angle=0,scale=.50]{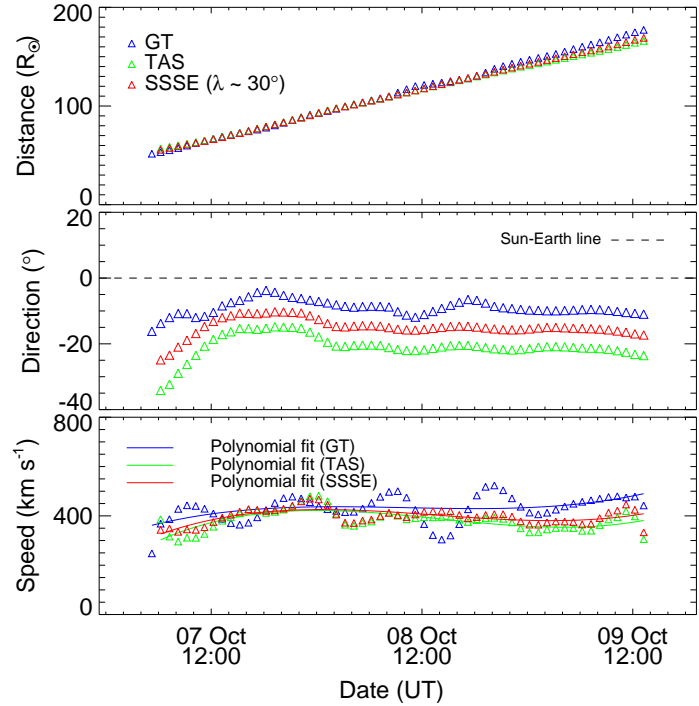}
\caption{\scriptsize{From top to bottom, panels show distance, propagation direction (relative to the Sun-Earth line) and speed profiles are shown for tracked CME feature as derived using the stereoscopic GT, TAS and SSSE methods. The horizontal line in the middle panel marks the Sun-Earth line.}}
\label{STAABB06Oct}
\end{center}
\end{figure}

\subsubsubsection{Geometric Triangulation (GT) Method using SECCHI/HI observations}
The use of stereoscopic HI (combined with COR2) observations to estimate the kinematics of CMEs was pioneered by \citet{Liu2010, Liu2010.722}, who proposed the Geometrical Triangulation (GT) technique. This technique assumes that the same point-like feature of an outward-moving CME can be tracked from the two STEREO viewpoints simultaneously. Neglecting projection effects (which can result in the two spacecraft observing different structures) and Thomson scattering effects, which is valid for Earth-directed events, the GT technique is based on the assumption that the difference in the elongation angle measured simultaneously from two viewpoints is due solely to the different viewing directions. This triangulation technique has been applied to CMEs at different STEREO spacecraft separation angles in studies that relate the imaging observations to near-Earth in situ measurements \citep{Liu2010, Liu2011, Liu2012, Liu2013, Mostl2010, Harrison2012, Temmer2012, Mishra2013}. The elongation angle profiles for the October 6 CME,  extracted from the STEREO-A and STEREO-B ecliptic J-maps, were interpolated onto a common time grid. We implemented the appropriate triangulation equations from \citet{Liu2010.722} to obtain the CME's kinematics. The derived distance, propagation direction and speed profiles of the CME (the latter derived from adjacent distance points) are shown in Figure~\ref{STAABB06Oct} in blue. The kinematic parameters obtained using the GT method at the sunward edge of the HI1 FOV are not shown in Figure~\ref{STAABB06Oct} due to occurrence of a singularity at those elongations (see \citealp{Liu2010, Mishra2013}).

We used the  kinematics derived using the GT method as input to the DBM to predict the CME arrival time and speed at L1. To initiate the DBM, we used a CME speed of 470 km s$^{-1}$ (the average of the last few values) at a distance of 177 R$_\sun$ (0.82 AU) at 13:14 UT on 2010 October 9. As before, the ambient solar wind speed was set to 350 km s$^{-1}$. The resultant L1 arrival times and speeds, corresponding to the extreme range of the drag parameter, are given in Table 1.

\subsubsubsection{Tangent to A Sphere (TAS) method using SECCHI/HI observations}
Soon after the development of GT by \citet{Liu2010}, \citet{Lugaz2010.apj} proposed a method for stereoscopic reconstruction of CMEs based on the Harmonic Mean (HM; \citealt{Lugaz2009}) geometry. This stereoscopic technique is referred to as the Tangent to A Sphere (TAS) method. We apply equation (2) of \citet{Lugaz2010.apj} to estimate the propagation direction of the tracked CME using this technique. As in the previous section, we use the derived time-direction profiles to estimate the distance, and hence the speed, profiles of the CME based on the expression for the radial distance of the transient's apex. The results are shown in Figure~\ref{STAABB06Oct} in green.

The central panel of Figure~\ref{STAABB06Oct} (green line) suggests that the CME is propagating slightly eastward of the Sun-Earth line. At the last point of its track (13:14 UT on October 9), the estimated CME distance, and speed are 166 R$_\sun$ (0.77 AU) and 385 km s$^{-1}$, respectively. The kinematics at the sunward edge of the HI1 FOV are not shown due to the occurrence of a singularity, as in the GT method. We applied the DBM exactly as discussed earlier; results are shown in Table 1.

\subsubsubsection{Stereoscopic Self-Similar Expansion (SSSE) method using SECCHI/HI observations}
Both GT and TAS methods described in the present section are based on extreme geometrical descriptions of solar wind transients (a point source for GT and an expanding circle attached to the Sun for TAS). Therefore, \citet{Davies2013} proposed a revised  technique based on the more generalized SSE geometry of \citet{Lugaz2010.apj} and \citet{Davies2012}. They named this the Stereoscopic Self-Similar Expansion (SSSE) method. They showed that the GT and TAS methods can be considered as the limiting cases  of the SSSE technique. We used the equations (23), (24) and (4a) of \citet{Davies2013} to estimate the propagation direction, from the observer, and distance profiles for the tracked CME. Although, this SSSE method can take any value of the half angular width ($\lambda$) of the CME between 0$\arcdeg$ and 90$\arcdeg$, we use a fixed value 30$\arcdeg$. The obtained kinematics for the tracked feature are  shown in Figure~\ref{STAABB06Oct} in red. The kinematics for the points at the sun-ward edge of HI FOV are not shown due to large errors. For these points, the sum of the elongation from both observers and the separation angle between the two observers is close to 180$\arcdeg$. As pointed out by \citet{Davies2013}, in such a situation, small errors in elongation will result in large errors in direction, and hence in distance and speed, around the aforementioned singularity. The estimated kinematics at the end of the track are used as inputs in the DBM to predict the arrival times and speeds at L1 (Table 1).

\subsection{2010 April 3 CME}
A geo-effective (D$_{st}$ = -72 nT) CME induced by a filament eruption associated with NOAA Active Region (AR) 11059 was observed as a halo by SOHO/LASCO C2 at 10:33 UT on April 3. SECCHI/COR1-A and B first observed this CME at 09:05 UT, in the SE and SW quadrants, respectively. We used the tie-pointing method of 3D reconstruction (scc$\_$measure: \citealt{Thompson2009}) on a selected feature along the leading edge of this CME and obtained its 3D kinematics. The 3D speed, latitude and longitude were estimated as 816 km s$^{-1}$, 25$\arcdeg$ South and 5$\arcdeg$ East of the Earth, respectively, at the outer edge of the COR2 FOV. The kinematics of this fast CME seem to be partly influenced by the presence of high speed solar wind, as the CME experiences little deceleration during its journey from the Sun to 1 AU; during this time the Earth was found to be immersed in fast solar wind. The kinematics of this CME have been studied extensively by previous authors  \citep{Mostl2010,Wood2011,Liu2011,Mishra2013}. In contrast to the 2010 October 6 event, this gives us an opportunity to assess the accuracy of various methods for the case of a fast CME moving in a fast ambient solar wind.

\begin{figure}
\begin{center}
\includegraphics[angle=0,scale=.50]{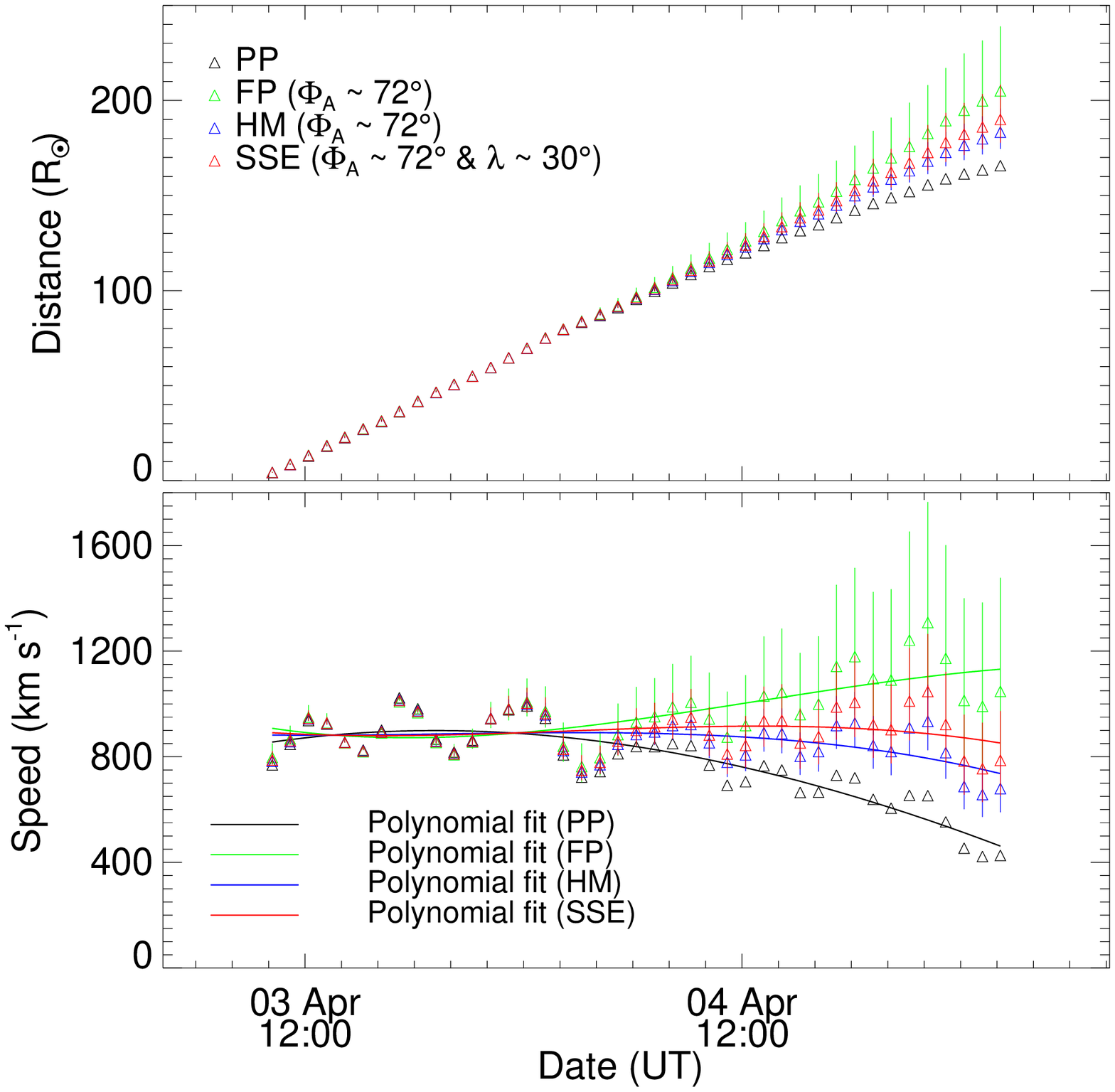}
\caption{\scriptsize{As Figure 3, for the 2010 April 3 CME.}}
\label{STAA03April}
\end{center}
\end{figure}

\begin{figure}
\begin{center}
\includegraphics[angle=0,scale=.50]{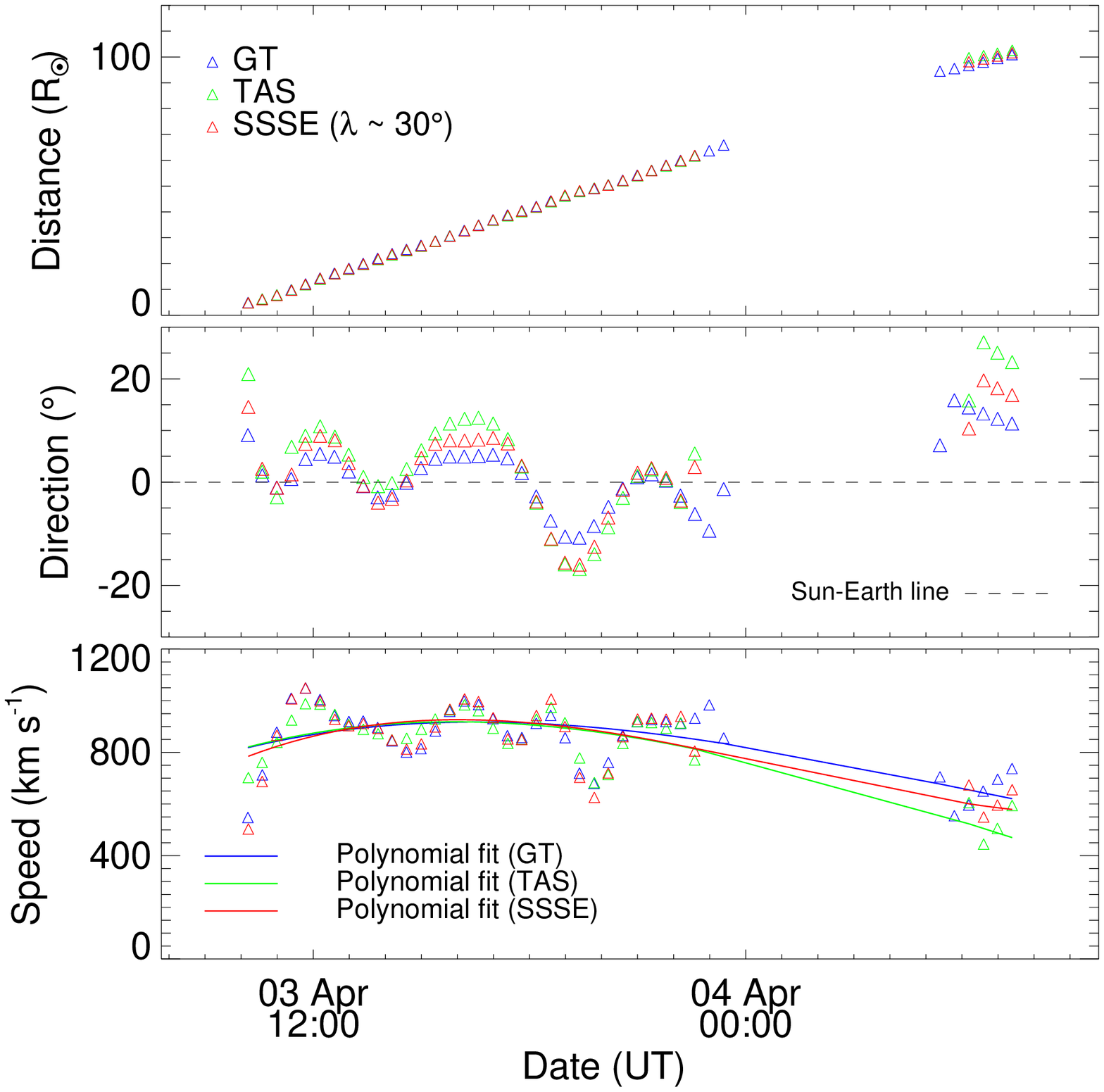}
\caption{\scriptsize{As Figure 5, for the 2010 April 3 CME.}}
\label{STAABB03April}
\end{center}
\end{figure}

We constructed STEREO-A and B ecliptic J-maps for this interval encompassing this CME using COR2 and HI images and extracted, from each, the time-elongation profile for the leading edge of the initial CME front. The Milky Way is visible in the HI2-B images, therefore the CME signal is less easily tracked. This CME can be tracked out to 54.5$\arcdeg$ and 26.5$\arcdeg$ elongation in STEREO-A and B J-maps, respectively. We implemented the seven single spacecraft methods (PP, FP, FPF, HM, HMF, SSE and SSEF) and the three stereoscopic methods (GT, TAS and SSSE) to derive the CME kinematics. Estimated kinematics from the PP, FP, HM and SSE methods, applied to the time-elongation profile of the CME extracted from the STEREO-A J-map, are shown in Figure~\ref{STAA03April}. Errors bars are calculated in the same manner as for the 2010 October 6 CME discussed in section 2.1.1. We also estimated the CME's kinematic properties based on the STEREO-B J-map. We used the kinematics from both spacecraft as inputs to the DBM to obtain the arrival time at L1 (given in Table 2). The kinematics of the tracked CME obtained using the stereoscopic methods are shown in Figure~\ref{STAABB03April}. Again, results for the arrival time and speed at L1, based on the use of these kinematics as input to the DBM model, are included in Table 2. Also results from FPF, HMF and SSEF analysis are quoted (corrected for off-axis propagation for the HMF and SSEF cases). Results from equivalent single spacecraft fitting analyses of STEREO-B data are also included. For this CME, we use an ambient solar wind speed 550 km s$^{-1}$ in the DBM. Note that only the minimum value of the statistical range of the drag parameter is used because the fast ambient solar wind into which this CME was launched is characterized by a low density.

\subsection{2010 February 12 CME} 
\begin{figure}
\begin{center}
\includegraphics[angle=0,scale=.45]{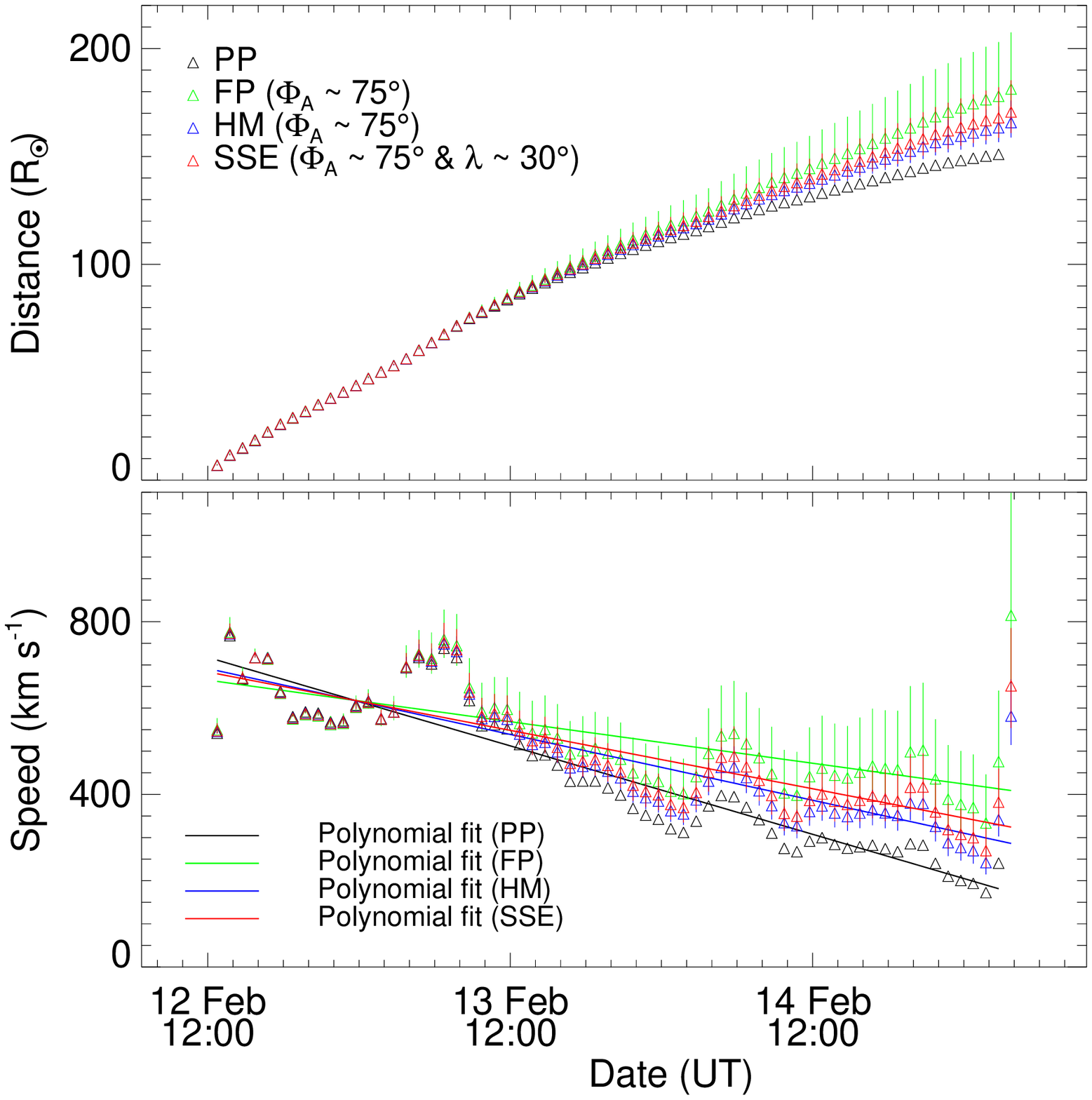}
\caption{\scriptsize{As Figure 3, for the 2010 February 12 CME.}}
\label{STAA12February}
\end{center}
\end{figure}

\begin{figure}
\begin{center}
\includegraphics[angle=0,scale=.45]{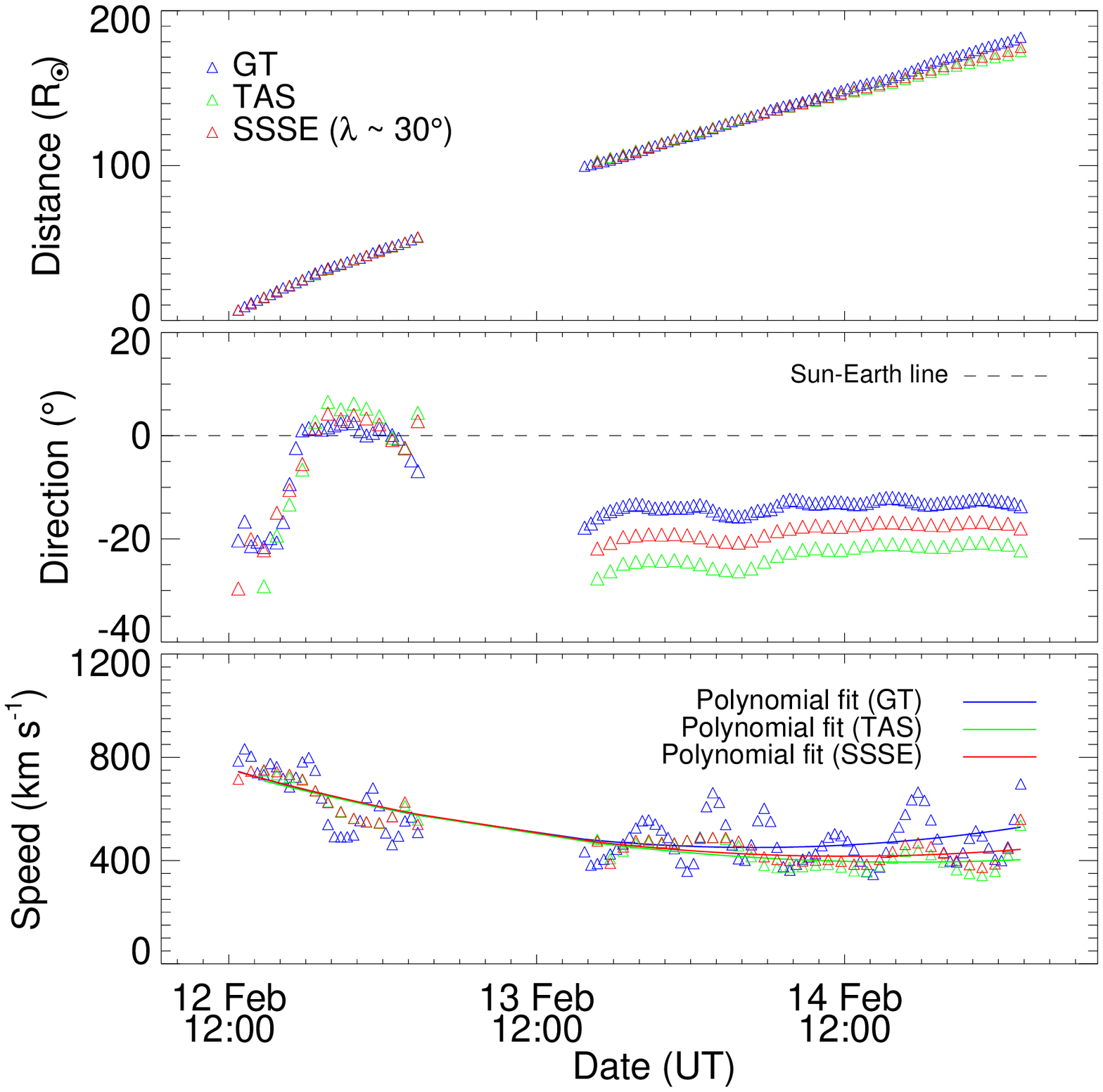}
\caption{\scriptsize{As Figure 5, for the 2010 February 12 CME.}}
\label{STAABB12February}
\end{center}
\end{figure}

This CME was first observed, as a halo, by SOHO/LASCO C2 at 13:42 UT on February 12. SECCHI/COR1 A and B first observed this CME at 11:50 UT in the NE and NW quadrants, respectively. We carried out 3D reconstruction of a selected feature along the leading edge of this CME using the scc$\_$measure procedure \citep{Thompson2009}, from which the 3D speed, latitude and longitude of the CME at the outer edge of COR2 FOV were estimated to be 867 km s$^{-1}$, 5$\arcdeg$ North and 10$\arcdeg$ East of the Earth, respectively. The heliospheric kinematics of this geo-effective CME (D$_{st}$ = -58 nT), and its near Earth in situ signatures, were studied by \citet{Mishra2013}, who showed that this fast CME continuously decelerated throughout its journey beyond the COR2 FOV to 1 AU. This CME provides us with an opportunity to test the efficacy of various methods to predict the arrival time of a fast CME decelerating in a slow ambient solar wind. We tracked this CME out to 48$\arcdeg$ and 53$\arcdeg$ in the STEREO-A and B J-maps constructed using COR2 and HI observations, respectively. We apply all of the analysis methods to this  CME in the same way as we did for the October 06 CME (Section 2.1). Results of the PP, FP, HM and SSE methods, applied to the time-elongation profile extracted from the STEREO-A J-map are shown in Figure~\ref{STAA12February}. Errors in distance and speed (marked  with  vertical  lines) are calculated in the same manner as for the previous CMEs. Results of the stereoscopic GT, TAS and SSSE techniques are shown in Figure~\ref{STAABB12February}. Again, for the latter techniques, we do not show points near the singularity. The estimated kinematics used as input in the DBM, and the resultant predicted arrival times and transit speeds, are given in Table 3. Results of the single-spacecraft fitting methods (FPF, HMF, and SSEF), applied to elongation profiles from both STEREO-A and STEREO B, are quoted in the bottom panel of Table 3.

\section{Identification of tracked CME features using in situ observations near Earth}

Classically, a CME imaged near the Sun shows a characteristic three part structure i.e. a bright leading edge followed by a dark cavity and finally a bright core \citep{Illing1985}. It is believed that leading edge, which appears bright due to the sweeping up of coronal plasma by erupting flux ropes or the presence of pre-existing material in the overlying fields \citep{Riley2008}, is identified near the Earth in the in situ observations as the CME sheath region (the disturbed region in front of the leading edge of the CME) \citep{Forsyth2006}. The darker region is assumed to correspond to a flux rope structure having a large magnetic field and a low plasma density, and is identified as a magnetic cloud (or MC) \citep{Klein1982} in in situ observations \citep{Burlaga1991}. In a classical sense, an MC is a plasma and magnetic field structure that shows an enhanced magnetic field, a rotation in magnetic field vector, a low plasma density and temperature, and a plasma $\beta$ of less than unity  \citep{Burlaga1981,Lepping1990,Zurbuchen2006}. The inner-most bright feature (the CME core) has been observed in H-$\alpha$ which indicates its cooler temperature. In the past, it has often been difficult to associate features imaged near the Sun with features observed in situ, due to the large distance gap and the difficulty  in characterizing the true evolution of the remotely-sensed features. However, using heliospheric imaging observations, it is now possible to relate the near Sun and near Earth observations of CMEs.

\begin{figure}
\begin{center}
\includegraphics[angle=0,scale=.70]{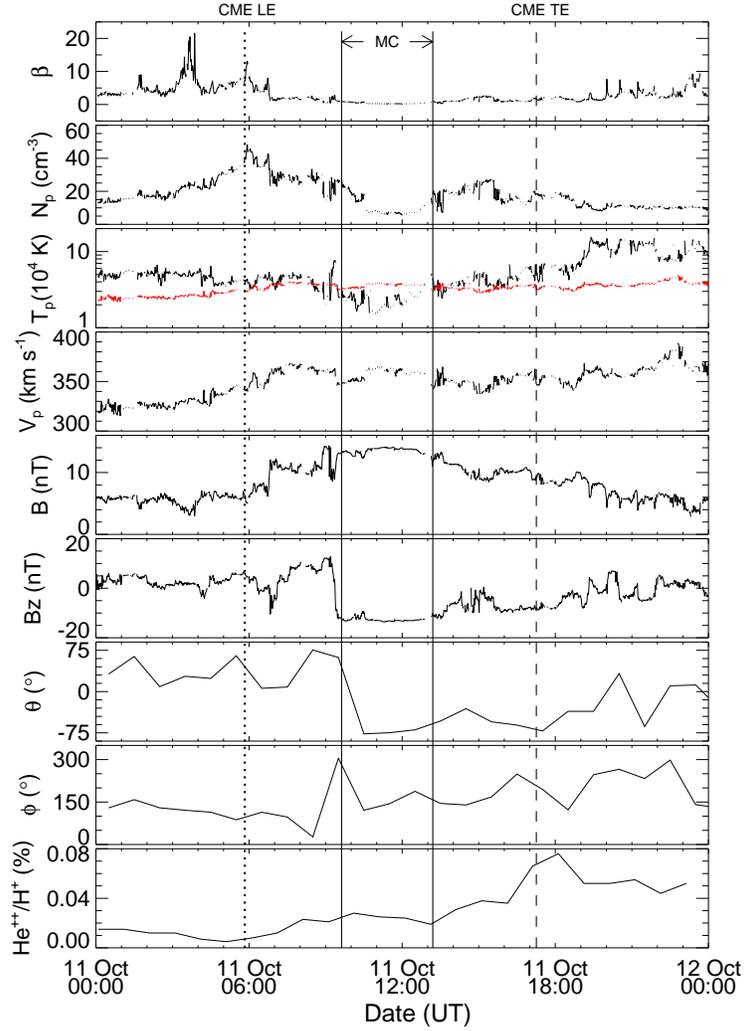}
\caption{\scriptsize{From top to bottom, panels show the in situ measurements at L1 of plasma beta, proton density, proton temperature, flow speed, magnetic field magnitude, magnetic field z-component, latitude and longitude of magnetic field vector, and the alpha to proton ratio. The red curve in the third panel shows the expected variation of the temperature.  From the left, the first (LE), second, third and fourth (TE) vertical lines mark the arrival of CME  leading edge, the leading edge of the magnetic cloud, the trailing edge of the magnetic cloud, and the trailing edge of the CME respectively.
}}
\label{Insitu}
\end{center}
\end{figure}

We have tracked the leading edge of the initial intensity front of the Earth-directed 2010 October 6 CME, in J-maps derived from STEREO/HI images, and derived its kinematics based on a number of techniques. We have also identified this CME in the in situ data taken by the ACE \citep{Stone1998} and Wind \citep{Ogilvie1995} satellites, based on plasma, magnetic field and compositional signatures \citep{Zurbuchen2006}. The in situ observations from 2010 October 11 are shown in Figure~\ref{Insitu}. The first vertical line in Figure~\ref{Insitu} (dotted, labeled LE) marks the arrival of the CME leading edge at the Wind spacecraft at 05:50 UT on 2010 October 11, and the fourth vertical line (dashed, TE) marks the trailing edge arrival at Wind at 17:16 UT. The region bounded by the second and third vertical lines (solid), at 09:38 UT and 13:12 UT respectively, can be classified as a magnetic cloud, as it shows an enhanced magnetic field ($>$10 nT), a decreased plasma $\beta$ ($<$ 1), and a smooth rotation in the magnetic field over a large angle ($>$ 30$\arcdeg$) \citep{Klein1982, Lepping1990}.

The feature of this CME tracked in the J-maps (Figure~\ref{Jmaps}) corresponds to the leading edge of the initial, curved CME-associated intensity front (Figure~\ref{Evolution}). Hence, its arrival is likely associated with LE. Thus 05:50 UT on 2010 October 11 is considered as the actual arrival time of the remotely sensed feature. Furthermore, the in situ speed of the CME is approximately 355 km s$^{-1}$ at L1. Tables 1 summarizes the differences between the range of predicted and actual (in situ at L1) arrival times and speeds for each method.

We also identified the 2010 April 03 CME in the in situ data taken at L1. The arrival times of the shock and the CME leading and trailing boundaries, based on plasma and magnetic field signatures are marked in Figure 14 of \citet{Mishra2013}. The arrival time and speed of the in situ signature, which is thought to be associated with the feature tracked in the STEREO/HI data, are 12:00 UT on 2010 April 05 and 720 km s$^{-1}$, respectively. The in situ arrival time and speed are used to compute errors in the predicted values from each methods (Table 2).

We also identified the 2010 February 12 CME in the near-Earth in situ data (see Figure 10 of \citealt{Mishra2013}). The L1 arrival time of this CME is considered to be 23:15 UT on 2010 February 15, and its speed 320 km s$^{-1}$. These values are used as a reference, to compute the errors in the predicted arrival times and speeds at L1 given in Table 3.

\section{Results and Discussions}
We constructed J-maps using SECCHI/HI and COR2 images (including the latter for only two of the CMEs) to extract time-elongation profiles for three selected CMEs in order to analyze their kinematic properties. We implemented four single spacecraft methods (PP, FP, HM and SSE), three single spacecraft fitting methods (FPF, HMF, and SSEF) and three stereoscopic methods (GT, TAS and SSSE).

For the CMEs of October 6 and February 12, the arrival time and speed predictions are slightly more accurate for all methods if the maximum value of the drag parameter is used in the DBM (see Tables 1 and 3). This is possibly due to fact that these CMEs are less massive, have a large angular width and are propagating in a dense solar wind environment \citep{Vrsnak2013}. From our study of three CMEs based on the 10 aforementioned techniques, we find that there are large errors involved in estimating their kinematic properties (up to 100 km s$^{-1}$ in speed) using HI data. It may be preferable to use 3D speeds determined in the COR2 FOV in the Sun-Earth direction for reliable and advance space weather forecasting, particularly for slow CMEs propagating in slow solar wind.

It is worth noting that the major contributions to arrival time errors arise due to limitations of the methods themselves. However, implementation of the DBM may also contribute to these errors. It is also important to point out that the selected CMEs in our study propagate within $\pm$ 20$\arcdeg$ of the Sun-Earth line, so, strictly, an off-axis correction is also required for the HM, SSE, TAS and SSSE methods before using the resultant speed as input to the DBM. However for these CMEs, we estimate that such a correction would decrease the speed by only a few km s$^{-1}$ and hence increase the predicted arrival time by only a few tens of minutes. We expect that if the final estimated speed of each tracked CME was taken as constant for the rest of the CME's journey to L1, then the errors in predicted arrival time would be similar to that obtained from using the DBM with the minimum drag parameter. This is because the CMEs are tracked out to a large fraction of 1 AU ($\approx$ 0.5 to 0.8 AU), and the small drag force applied only beyond that distance will have little effect on the CME dynamics.

Our assessment of the relative performance of various methods is based mainly on the difference between the CME arrival time predicted at L1 by those methods and the ``actual'' arrival time determined in situ. We do not perform a detailed comparison of  the kinematic profiles derived using each method. Of course, different kinematics profiles can lead to the same arrival time.  Moreover, it is unlikely that same part of a CME that is being tracked in remote imaging observations will pass over the spacecraft that is making the single-point in situ measurements. We cannot advocate, with confidence, the superiority of one method over others based only on it producing even the most accurate of arrival time predictions. It would be useful to compare the speed profile of a CME derived in the inner heliosphere using 3D  MHD modeling  with  CME dynamics  derived using the mainly-HI driven techniques applied in the current study. Of course, results from MHD models also need to be considered with caution.

\subsection*{Relative performance of single spacecraft techniques}

In our study, we assess the performance of various reconstruction techniques based on the obtained difference between the predicted and actual arrival time of three CMEs. Of the four single spacecraft methods that do not rely on a curve fitting approach, the Point-P (PP) method gives the largest range of errors (up to 25 hours) in predicted arrival time over all CMEs. This is perhaps due to its oversimplified geometry. At large elongations (beyond $\approx$ 120 R$_\sun$), for the CMEs of October 6 and February 12, the speed estimated using the PP technique is less than the ambient solar wind speed, which is unphysical. For the October 6 CME, the PP, FP, HM and SSE approaches produce roughly similar errors in predicted arrival time (up to 25 hours) and transit speed (up to 100 km s$^{-1}$). For the CME of April 3, the HM and PP methods provide the most and least accurate predictions of arrival time at L1, respectively. For the CME of February 12, the SSE method provides the most accurate L1 arrival time while the PP method is the least accurate of these four methods.

The kinematics of the April 3 CME estimated using the FP technique (Figure~\ref{STAA03April}: green trace) show a sudden unphysical late acceleration,  possibly due to its real deflection. From this, we suggest that the FP, HM and SSE methods, as implemented here, can give more accurate results if the estimated speed tends to a constant value far from the Sun. In the FP method, the tracked feature is assumed to correspond to the same point moving in a fixed radial direction, which is unlikely to be valid for a real CME structure \citep{Howard2011}. One major drawback of the FP method is that it does not take into account the finite cross-sectional extent of a CME. In terms of the four single viewpoint methods that enable estimation of the kinematics properties as a function of time, the HM and SSE methods provide a more accurate arrival time prediction for this CME. Of course, the assumption of a circular front in these methods may not be totally valid due to possible flattening of the CME front resulting from its interaction with the structured coronal magnetic field and solar wind ahead of the CME \citep{Odstrcil2005}. Also, the assumption made here in implementing the HM and SSE methods (and indeed the FP method) that the CME propagates along a fixed radial trajectory (in particular one derived close to the Sun), ignoring real or ``artificial'' heliospheric deflections, will induce errors, particularly for slow speed CMEs that are more likely to undergo deflection in the IP medium \citep{Wang2004,Gui2011}. As noted previously, by ``artificial deflection'' we mean that the observer is not detecting the same feature of CME in consecutive images; this is due to well-known geometrical effect. We conclude that the implausible deceleration of October 6 CME, to a speed less than that of the ambient solar wind speed (see Figure~\ref{STAA06Oct}), is due to violations of the assumptions inherent in the PP, HM and SSE methods at elongations beyond 30$\arcdeg$.

Irrespective of event, the PP, FP, HM and SSE methods estimate significantly different radial distance and speed profiles after approximately 100 R$_{\sun}$. This is because the assumed geometry has more impact on the results with increasing elongation. For the PP and FP methods, implausible acceleration or deceleration evident beyond approximately 100 R$_{\sun}$, if assumed to be real, would lead to unrealistically large errors in arrival time prediction. Therefore, among the single spacecraft methods, we suggest that methods like HM and SSE should be used to achieve reasonable arrival time predictions.

The value of the propagation direction that is adopted for each CME in our implementation of the FP, HM and SSE techniques will affect the performance of each method; this is an important issue in our study.  The  quoted  CME arrival  times  in Tables  1, 2 and 3 based on the FP, HM and  SSE techniques are based on a direction estimated from tie-pointing. To assess the sensitivity of our results to the exact value of the propagation direction used, we have repeated our analysis using a range of propagation directions.  As described in Section 2.1.1, we repeated our FP, HM and SSE analyses using propagation directions that are +10$\arcdeg$ and -10$\arcdeg$ different to the values ($\phi$) estimated using tie-pointing.  We used these revised kinematic profiles to estimate the arrival time and transit speed of the selected CMEs in our study for $\phi$ $\pm$ 10$\arcdeg$.

For the October 6 CME analyzed using the FP method, using $\phi$+10 ($\phi$-10) resulted in predicted arrival times that are 21 hours later (earlier) for STEREO-A and 10 hours earlier (later) for STEREO-B than the arrival time predicted using $\phi$.  For the HM method, using $\phi$+10 ($\phi$-10) resulted in a predicted arrival time 12 hours later (earlier) for STEREO-A and 6 hours earlier (later) for STEREO-B. For the SSE method,  using $\phi$+10 ($\phi$-10) results in a predicted arrival time 16 hours  later (earlier) for STEREO-A and 7 hours earlier (later) for STEREO-B. For the 2010 April 3 CME, the deviation in arrival time  from that quoted in Table 2 is less than 5 hours for STEREO-A and less than 3 hours for STEREO-B for all the three single spacecraft methods for  $\phi$+10. Assuming a propagation direction equal to $\phi$-10 yields the same uncertainties as for $\phi$+10, except for the FP  method applied to STEREO-A where the uncertainty increases to 8 hours. For the 2010 February 12 CME, assuming that the propagation direction is $\phi$+10 ($\phi$-10) in FP analysis results in predicted arrival times 9 hours later (14 hours earlier) for STEREO-A and 4 hours earlier (6 hours later) for STEREO-B than the values quoted in Table 3 that are based on using $\phi$ directly from tie-pointing. In case of the HM and SSE methods, using $\phi$+10 ($\phi$-10) results in predicted arrival times that are less than 9 hours later (earlier) for STEREO-A and less than 6 hours earlier (later) for STEREO-B than the values quoted in Table 3.

The uncertainties discussed above do not reflect the total errors involved in the implementation of the FP, HM and SSE methods. The uncertainties in the distance and speed due to a change in propagation direction are not significant at smaller elongations \citep{Wood2009,Howard2011}. Therefore, for a case like that of the April 3 CME as observed by STEREO-B, in particular,  where the CME cannot be tracked out far in elongation angle, any uncertainty in propagation direction will have a minimal effect on the derived kinematic profile and also the predicted arrival time. At greater elongations, however, this effect, along with the ``well-known effect of CME geometry'', will severely limit the accuracy of these  methods. Given the effects of uncertainty in propagation direction in the FP, HM and SSE methods, we note that it may be better to combine the DBM with CME kinematics derived in the near-Sun HI1 FOV to  optimize the goal of space weather prediction, at least for CMEs launched into a slow speed ambient solar wind medium.

\subsection*{Relative performance of stereoscopic techniques}

For the October 6 CME, the stereoscopic GT method gives a large range of arrival time errors (up to 25 hours) while the stereoscopic TAS method predicts arrival time to within 17 hours. For this CME, the errors resulting from the application of the SSSE method are intermediate between those from the GT and TAS techniques. For the February 12 CME, among the three stereoscopic methods the TAS method provides the best prediction of L1 arrival time (within 2 hours of the in situ arrival) and transit speed (within 45 km s$^{-1}$ of the in situ speed). For the April 3 CME, all of the stereoscopic methods give approximately the same  arrival time errors (within 8 hours).

As in the FP (and FPF) techniques, the assumption in GT that the same point of a CME is being observed in consecutive images, is likely to become increasingly invalid with increasing elongation. GT also assumes that the same point is observed simultaneously from both viewpoints. Moreover, the effect of ignoring the Thomson scattering geometry is minimized for Earth directed events, therefore deviations from such a configuration will also result in errors in the estimated kinematics. From our analysis, which is limited to three CMEs, we conclude that, among the three stereoscopic methods, TAS technique performs most accurately and GT performs least accurately in estimating CME arrival times and speeds. The SSSE arrival time is within a couple of hours of that from the TAS technique. As the estimated kinematics properties from the SSSE method (here implemented with an angular half width of 30$\arcdeg$) are intermediate between the kinematics derived from the GT and TAS methods (see Figure~\ref{STAABB06Oct}, \ref{STAABB03April}, \ref{STAABB12February}), we are tempted to suggest that the SSSE method may be preferable for space weather forecasting if a reasonable estimate of a CME's half angular width is available. In all of the stereoscopic methods used, any effects due to the Thomson scattering geometry are ignored and the assumption of self-similar expansion \citep{Xue2005} may also result in errors. However, one needs to quantify the potential errors due to ignoring real effects for each method, over different elongation ranges and for different spacecraft separation angles, before concluding the unbiased superiority of the TAS technique.

\subsection*{Relative performance of single spacecraft fitting techniques}

Results from the single spacecraft fitting techniques (FPF, HMF, and SSEF) suggest that the October 6 CME propagates eastward of the Sun-Earth line. All three methods give roughly the same launch time for the CME. Using STEREO-A observations, the error in the predicted CME arrival time at L1 is least (within 5 hours) for HMF and largest (within 30 hours) for FPF. Arrival time errors derived from STEREO-B observations are similar for all three methods ($\approx$ 22 hours). For the fast CME of 2010 April 3, which did not decelerate noticeably, arrival time errors are small (within 5 hours) for all three fitting methods. For this CME, SSEF predicts most accurately the arrival time at L1 while FPF is least accurate. Note that the longitudes output by these fitting methods are up to 40$\arcdeg$ from the Sun-Earth line for both the October 6 and April 3 CMEs. We consider these CMEs to be closely Earth directed. For the fast, decelerating CME of 2010 February 12, arrival time errors from these single spacecraft fitting techniques are very large (18 to 33 hours) and the estimated CME longitudes (Table 3) can be more than 70$\arcdeg$ from the Sun-Earth line. The large errors in the results from applying these fitting methods to slow or decelerating CMEs is most likely due to a breakdown in their inherent assumptions of constant speed and direction. The predicted arrival times, and errors therein, and errors in transit speed resulting from application of the SSEF technique to the STEREO-A profiles are not shown for the October 6 and April 3 CMEs. In these cases, the CME is not predicted to hit an in situ spacecraft at L1, based on retrieved propagation direction. Note that all of the fitting methods reproduce the observed elongation track well, so we must be cautious on relying on the fitted parameters to consider one method as superior.

In terms of the three fitting methods (FPF, HMF and SSEF), we find that HMF and SSEF (here applied with $\lambda$ = 30$\arcdeg$) more accurately predict arrival time and transit speed at L1 than does the FPF method. For all three CMEs, the propagation
direction derived from applying these fitting techniques separately to the STEREO-A and STEREO-B elongation variations show little consistency (see the Tables 1, 2 $\&$ 3). However, based on our study, it is suggested that CME propagation direction is best obtained using the FPF  method while it is worst from the HMF method.  \citet{Lugaz2010} has shown  that  the  FPF  method can  give  significant errors in propagation direction when a CME is propagating at an angle beyond 60$\arcdeg$ $\pm$ 20$\arcdeg$ from the Sun-spacecraft line; this is not in agreement with our findings. We also note that in case of the fast 2010 February 12 CME, where the assumption of a constant CME speed is not valid i.e.  a physical deceleration is observed, the propagation
direction estimated from all fitting methods are highly erroneous. These elongation profile fitting approaches have potential to give better results if features are tracked out to large elongations (at least 40$\arcdeg$) and the manual selection of points is done with extreme care \citep{Williams2009}.

\section{Conclusion}

From the application of ten methods to three Earth-directed CMEs observed by STEREO, having different speeds and launched into different ambient solar wind environments, we found that stereoscopic methods are more accurate than single spacecraft methods for the prediction of CME arrival times and speeds at L1. Irrespective of the characteristics of the CMEs, among the three stereoscopic methods the TAS method gives the best prediction of transit speed (within few tens of km s$^{-1}$) and arrival time (within 8 hours for fast CMEs and 17 hours for slow or fast decelerating CMEs).

We also find that the HM method (based on a propagation direction retrieved from 3D reconstruction of COR2 data) performs best among the single spacecraft techniques that we applied. For our fast speed CME with no apparent deceleration, the HM method provides the best estimate of the predicted L1 arrival time (within 2 hours) and speed (within 60 km s$^{-1}$). However, for our  fast but decelerating CME, this method predicts arrival time to around 10 hours, and this increases to $\approx$ 20 hours for the slow CME in our study.

Independent of the characteristics of the CME, our study shows that, among the techniques that we used, the HMF and SSEF single spacecraft fitting methods perform better than FPF. All three fitting methods give reasonable arrival time predictions (within 5 hours of the arrival time identified in situ) for the fast speed CME that undergoes no discernable deceleration. For the slow CME and the fast but decelerating CME, the fitting methods are only accurate to 10 to 30 hours in terms of their arrival time prediction and yield relatively larger errors (up to hundreds of km s$^{-1}$) in predicted speed.

Propagating the CME out to L1, using the DBM, appears to reduce the errors in predicted arrival time and speed for the cases where the CME is launched into a slow solar wind (represented by the use of a high drag parameter). This is particularly true in cases where the HI instruments are able to track CMEs to moderate distances.

In summary, the HI imagers provide an opportunity for us to understand the association between remotely observed CME structures and in situ observations. Our study demonstrates the difficulties inherent in reliably predicting CME propagation direction, and arrival time and speed at 1 AU, based on such remote-sensing observations. From our study, we conclude that, although HIs provide the potential to improve the space weather forecasting, for slow or decelerating CMEs, there are specific assumptions in some of the currently-used techniques that compromise the estimates of CMEs kinematics, and hence predictions of arrival time at Earth.

\section*{Acknowledgments} We acknowledge the UK Solar System Data Centre for providing the processed Level 2 STEREO/HI data.  The in situ measurements of solar wind data from ACE and WIND spacecraft were obtained from NASA CdAweb $(http://cdaweb.gsfc.nasa.gov/)$. We acknowledge the use of Drag Based Model, developed by Bojan Vr$\breve{s}$nak and available at $http://oh.geof.unizg.hr/CADBM/cadbm.php$, in our study. The work by N.S. partially contributes to the research for European Union Seventh Framework Programme (FP7/2007-2013) for the Coronal Mass Ejections and Solar Energetic Particles (COMESEP) project under Grant Agreement No. 263252. We thank the anonymous referee for his/her detailed and constructive comments which improved this paper.

\clearpage


\clearpage

\begin{landscape}
\begin{table}
  \centering
{\scriptsize
 \begin{tabular}{ p{3.0cm}|p{2.0cm}| p{3.0cm}| p{2.5cm}|p{2.5cm}|p{2.5cm}}
    \hline
		
 Method & Kinematics as inputs in DBM [t$_{0}$, R$_{0}$ (R$_\sun$), v$_{0}$ (km s$^{-1})$]& Predicted arrival time using kinematics + DBM (UT) [$\gamma$ = 0.2 to 2.0 (10$^{-7}$ km$^{-1}$)] & Predicted transit speed at L1 (km s$^{-1}$)  [$\gamma$ = 0.2 to 2.0 (10$^{-7}$ km$^{-1}$)] & Error in predicted arrival time (hrs) [$\gamma$ = 0.2 to 2.0 (10$^{-7}$ km$^{-1}$)] &  Error in predicted speed (km s$^{-1}$)  [$\gamma$ = 0.2 to 2.0 (10$^{-7}$ km$^{-1}$)]  \\  \hline

PP (STEREO-A) & 09 Oct 09:14, 130, 180  & 12 Oct 07:29 to 11 Oct 14:54  & 259 to 328 	& 25.6 to 9.05 	& -96 to -28  \\ \hline

PP (STEREO-B) & 09 Oct 21:27, 158, 300  & 11 Oct 08:15 to 11 Oct 06:41  &   306 to 327  &  2.4 to 0.8  &  -50 to -28  \\  \hline

FP (STEREO-A) & 09 Oct 13:14, 174, 390  & 10 Oct 08:23 to 10 Oct 08:45	& 388 to 376 	& -21.4 to -21	& 33 to 21  \\ \hline

FP (STEREO-B) & 09 Oct 07:27, 162, 435 &  10 Oct 06:11 to 10 Oct 07:45 & 425 to 384  & -23.6 to -22.1  & 70 to 30 \\ \hline

HM (STEREO-A) 	& 09 Oct 13:14, 149, 230 & 11 Oct 14:22 to 11 Oct 06:21	& 266 to 324 	& 8.5 to 0.55	& -89 to -31   \\ \hline

HM (STEREO-B)  & 09 Oct 21:27, 189, 410 &  10 Oct 08:35 to 10 Oct 08:52 & 407 to 390  & -21.3 to -21 & 52 to 35   \\ \hline

SSE (STEREO-A) & 09 Oct 13:14, 157, 255  & 11 Oct 05:34 to 10 Oct 23:48	 & 276 to 322  &	0.1 to -6.0	 &  -79 to -33  \\ \hline

SSE (STEREO-B) & 09 Oct 21:27, 193, 470  & 10 Oct 05:33 to 10 Oct 06:03	 & 462 to 419  &	-24.3 to -23.8	&  107 to 64  \\ \hline

GT	& 09 Oct 13:14, 177, 470 &  10 Oct 04:04 to 10 Oct 05:24	& 456 to 400 	& -25.8 to -24.4	& 101 to 45 \\  \hline

TAS &  09 Oct 13:14, 166, 385 &  10 Oct 12:39 to 10 Oct 13:04	 & 383 to 372 	& -17.1 to -16.7 & 28 to 17    \\  \hline
SSSE &  09 Oct 13:14, 169, 410 &  10 Oct 09:53 to 10 Oct 10:41	& 405 to 381 	& -19.9 to -19.2	 & 50 to 26    \\  \hline		
			
 \end{tabular}

\begin{tabular}{p{3.0cm}| p{3.0cm}| p{2.5cm}|p{2.5cm}|p{2.5cm}| p{2.0cm}}
   
\multicolumn{6}{c}{Time-elongation track fitting methods} \\  \hline
   
 Methods   & Best fit parameters [t$_{(\alpha = 0)}$, $\Phi$ ($\arcdeg$), v (km s$^{-1}$)]  &   Predicted arrival time at L1 (UT)   & Error in predicted arrival time  & Error in predicted speed at L1 (km s$^{-1}$) & Longitude ($\arcdeg$) 
   \\ \hline
	
	FPF (STEREO-A)   & 06 Oct 07:38, 96.2, 462 & 10 Oct 00:39 & -29.1  &  107      &  -13   \\  \hline
	FPF  (STEREO-B)  & 06 Oct 05:05,61, 414  & 10 Oct 08:25  & -21.3  &    59     &  -17 \\  \hline
	 
	HMF (STEREO-A)   & 06 Oct 08:40, 136, 610 & 11 Oct 00:19 & -5.5  &    12     &   -53  \\  \hline
	HMF  (STEREO-B)  & 06 Oct 06:53, 74.5, 434 & 10 Oct 05:50 &  -24  &    78    &   -4 \\  \hline
	
	SSEF (STEREO-A)  & 06 Oct 08:16, 115, 525  & ---   &   --- &   --- &   -32  \\  \hline
	SSEF (STEREO-B)  & 06 Oct 06:19, 68.1, 426  & 10 Oct 09:57  & -19.8  &  55    &   -10   \\  \hline

\end{tabular}
}
\caption{\scriptsize{Results of the applied techniques for the 2010 October 6 CME (first column). 
Upper section -- Second column: the kinematics properties as output directly by the techniques and used as input  to the  DBM. Third and fourth columns: predicted  arrival time  and speed of this CME at  L1 corresponding  to the extreme range of the drag parameter used in the DBM.  Fifth and sixth columns: errors in predicted arrival time and speed based on comparison with in situ arrival time and speeds. 
Lower section (fitting techniques) -- Second column: best fit launch time, longitude from observer and speed. Third column: predicted arrival time at L1, Fourth and fifth columns: errors in predicted arrival time and speed (computed as above). Sixth column: longitude from Sun-Earth line. 
The STEREO-A and B shown in parentheses for each method denotes the spacecraft from which the derived elongation is used. Negative (positive) errors in predicted arrival time correspond to a predicted arrival time that is before (after) the actual CME  arrival time determined from in situ measurements.  Negative (positive) errors in predicted speed correspond to a predicted speed that is less (more)  than the actual speed of the CME  at L1.}}

\end{table}
\end{landscape}

\begin{landscape}
\begin{table}
  \centering
{\scriptsize
 \begin{tabular}{ p{3.0cm}|p{2.0cm}| p{3.0cm}| p{2.5cm}|p{2.5cm}|p{2.5cm}}
    \hline
		
 Method & Kinematics as inputs in DBM [t$_{0}$, R$_{0}$ (R$_\sun$), v$_{0}$ (km s$^{-1})$]& Predicted arrival time using kinematics + DBM (UT) [$\gamma$ = 0.2 (10$^{-7}$ km$^{-1}$)] & Predicted transit speed at L1 (km s$^{-1}$)  [$\gamma$ = 0.2 (10$^{-7}$ km$^{-1}$)] & Error in predicted arrival time (hrs) [$\gamma$ = 0.2 (10$^{-7}$ km$^{-1}$)] &  Error in predicted speed (km s$^{-1}$)  [$\gamma$ = 0.2 (10$^{-7}$ km$^{-1}$)]  \\  \hline

PP (STEREO-A) & 05 April 02:11, 165, 426  & 05 April 23:14 & 446 	& 11.2 	& -274  \\ \hline

PP (STEREO-B) & 04 April 05:00, 85, 866 & 05 April 12:14 &  735 &  0.2  &  15  \\  \hline

FP (STEREO-A) &  04 April 12:11, 122, 800  & 05 April 11:14 & 727 	& -0.7 & 7  \\ \hline

FP (STEREO-B) & 04 April 04:59, 85, 790  &  05 April 14:20 & 702  & 2.3  & -18\\ \hline

HM (STEREO-A) & 05 April 02:11, 183, 660  & 05 April 10:53  & 653 	& -1.1 & -67   \\ \hline

HM (STEREO-B)  & 04 April 05:00, 85, 810 &  05 April 13:44 & 711  & 1.7 &  -9   \\ \hline

SSE (STEREO-A) & 05 April 02:11, 190, 800  & 05 April 07:42  & 777 	& -4.3 & 57   \\ \hline

SSE (STEREO-B) & 04 April 05:00, 85, 820  & 05 April 13:27  & 716 	& 1.5 & -4   \\ \hline

GT	       & 04 April 07:23, 101, 640    &  05 April 17:35	& 624	&  5.5 & -96  \\  \hline

TAS       &  04 April 07:23, 103, 580  &  05 April 19:57	 & 578 	& 7.9 & -142    \\  \hline
SSSE       & 04 April 07:23, 101, 615  &  05 April 19:51	& 574 	& 7.8  &  -146    \\  \hline		
			
 \end{tabular}

\begin{tabular}{p{3.0cm}| p{3.0cm}| p{2.5cm}|p{2.5cm}|p{2.5cm}| p{2.0cm}}
   
\multicolumn{6}{c}{Time-elongation track fitting methods} \\  \hline
   
 Methods   & Best fit parameters [t$_{(\alpha = 0)}$, $\Phi$ ($\arcdeg$), v (km s$^{-1}$)]  &   Predicted arrival time at L1 (UT)  & Error in predicted arrival time  & Error in predicted speed at L1 (km s$^{-1}$) & Longitude ($\arcdeg$) 
   \\ \hline
	
	FPF (STEREO-A)   & 03 April 08:47, 63.4, 865 & 05 April 08:18 &  -3.7 &  145      &  4  \\  \hline
	FPF  (STEREO-B)  & 03 April 09:07, 86.8, 886 & 05 April 07:30 & -4.5  &   166     &  16    \\  \hline
	 
	HMF (STEREO-A)   & 03 April 09:22, 78.5, 908  & 05 April 07:32 & -4.5  &   171      &   -11  \\  \hline
	HMF  (STEREO-B)  & 03 April 09:11, 103.5, 928 & 05 April 13:38 &  1.5  &    67    &   32 \\  \hline
	
	SSEF (STEREO-A)  & 03 April 09:11, 71, 889 & 05 April 07:39   & -4.3  &   163  &  -4  \\  \hline
	SSEF (STEREO-B)  & 03 April 09:10, 95, 907  & 05 April 17:29  & 5.5  &  12  &   24    \\  \hline

\end{tabular}
}
\caption{\scriptsize{The predicted arrival times and speeds (and errors therein) at L1 for the 2010 April 03 CME. Details as in the caption of Table 1.}}

\end{table}
\end{landscape}

\begin{landscape}
\begin{table}
  \centering
{\scriptsize
 \begin{tabular}{p{3.0cm}|p{2.0cm}| p{3.0cm}| p{2.5cm}|p{2.5cm}|p{2.5cm}}
    \hline
		
 Method & Kinematics as inputs in DBM [t$_{0}$, R$_{0}$ (R$_\sun$), v$_{0}$ (km s$^{-1})$] &  Predicted arrival time using kinematics + DBM (UT) [$\gamma$ = 0.2 to 2.0 (10$^{-7}$ km$^{-1}$)] & Predicted transit speed at L1 (km s$^{-1}$)  [$\gamma$ = 0.2 to 2.0 (10$^{-7}$ km$^{-1}$)] & Error in predicted arrival time (hrs) [$\gamma$ = 0.2 to 2.0 (10$^{-7}$ km$^{-1}$)] &  Error in predicted transit speed (km s$^{-1}$)  [$\gamma$ = 0.2 to 2.0 (10$^{-7}$ km$^{-1}$)] \\  \hline

PP (STEREO-A)	& 15 Feb 03:44, 152, 270  & 16 Feb 21:46 to 16 Feb 17:46	 &  286 to 325 & 	22.4 to 18.4 &  -34 to 5  \\ \hline

PP (STEREO-B)	& 15 Feb 01:43, 172, 330  & 16 Feb 01:26  to 16 Feb 01:16 &  331 to 335   & 	2.2 to 2	 &  11 to 15\\ \hline
  
FP (STEREO-A)	& 15 Feb 03:44, 181, 400 &  15 Feb 19:02 to 15 Feb 19:23  & 	397 to 382 &	-4.3 to -4.8	& 77 to 62 \\ \hline

FP (STEREO-B)	& 15 Feb 01:43, 187, 500 & 15 Feb 11:45 to 15 Feb 12:43 & 	485 to 419 &	-11.5 to -10.2	& 165 to 99  \\ \hline
  
HM (STEREO-A)	& 15 Feb 03:44, 165, 300 & 16 Feb 10:06 to 16 Feb 08:50 & 305 to 326 &	10.8 to 9.5 &  -15 to 6 \\ \hline

HM (STEREO-B)	& 15 Feb 01:43, 179, 420   &  15 Feb 17:15 to 15 Feb 17:52 & 415 to 389  	 & -6 to -5.5 &  95 to 69 \\  \hline

SSE (STEREO-A)	& 15 Feb 03:44, 171, 390 & 16 Feb 00:23 to 16 Feb 00:48 & 388 to 375 &	1.1 to 1.5 &  68 to 55 \\ \hline

SSE (STEREO-B)	& 15 Feb 01:43, 182, 450 & 15 Feb 14:58 to 15 Feb 15:48 & 441 to 400 &	-8.2 to -7.4 &  121 to 80 \\ \hline

GT         &  15 Feb 01:43, 183, 450  &  15 Feb 14:35 to 15 Feb 15:20 & 442 to 401  &  -8.7 to -7.9 & 122 to 81  \\  \hline

TAS       &  15 Feb 01:43,  174, 365  &  15 Feb 22:08 to 15 Feb 22:12 & 365 to 362 &   -1.1 to -1.0 &  45 to 42  \\  \hline
SSSE      & 15 Feb 01:43, 176, 400   &   15 Feb 19:12 to 15 Feb 19:39	& 397 to 380 	&  -4.0 to -3.6	 & 77 to 60   \\  \hline		

 \end{tabular}

\begin{tabular}{p{3.0cm}| p{3.0cm}| p{2.5cm}|p{2.5cm}|p{2.5cm}| p{2.0cm}}
\multicolumn{6}{c}{Time-elongation track fitting methods} \\  \hline
  
 Methods   & Best fit parameters [t$_{(\alpha = 0)}$, $\Phi$ ($\arcdeg$), v (km s$^{-1}$)]  &   Predicted arrival time at L1 (UT)   & Error in predicted arrival time  & Error in predicted speed at L1 (km s$^{-1}$) & Longitude ($\arcdeg$) 
   \\ \hline
	
	FPF (STEREO-A)   & 12 Feb 10:47, 93, 710 & 14 Feb 20:43 & -26.5  &  390     &  -28   \\  \hline
	FPF  (STEREO-B)  & 12 Feb 10:34, 77.7, 667 & 15 Feb 00:10  & -23  &   347   &  7 \\  \hline
	 
	HMF (STEREO-A)   & 12 Feb 11:18, 132, 926 & 17 Feb 08:16 &  33   &    41    &  -67  \\  \hline
	HMF  (STEREO-B)  & 12 Feb 11:19, 105.7, 764  & 15 Feb 04:46  & -18.5  &   305    &   35 \\  \hline
	
	SSEF (STEREO-A)  & 12 Feb 11:07, 111.7, 803 &  ---   &  ---   &   ---     &     -47   \\  \hline
	SSEF (STEREO-B)  & 12 Feb 11:04, 91.3, 714  & 15 Feb 05:37  &  -17.6  &  300   &     20 \\  \hline

\end{tabular}
}
\caption{\scriptsize{The predicted arrival times and speeds (and errors therein) at L1 for the 2010 February 12 CME. Details as in the caption of Table 1.}}
\end{table}

\end{landscape}

\clearpage

\end{document}